\documentclass[prd, twocolumn, nofootinbib, reprint, preprintnumbers, showkeys, floatfix]{revtex4-1}
\usepackage{color}
\usepackage{amsmath,amsfonts,bm}
\usepackage{graphicx, epstopdf,scrextend}
\usepackage{mathrsfs,mathtools}

\usepackage{pgfplots,pgfplotstable}
\usepgfplotslibrary{fillbetween}
\usepgfplotslibrary{groupplots}
\usetikzlibrary{intersections}
\pgfplotsset{compat=newest}

\pgfplotsset{every axis/.style={
    width=8.6cm,
    height=8.6cm,
    grid=both,
    scaled ticks=false,
    yticklabel style={/pgf/number format/.cd, fixed,precision=5}
  }
}

\definecolor{darkgreen}{rgb}{0,0.4,0} 
\definecolor{darkblue}{rgb}{0,0,0.6} 
\usepackage[colorlinks=true,linkcolor=darkblue,citecolor=darkgreen]{hyperref}

\newcommand{\as}{\alpha_{\mathrm{s}}}

\newcommand{\LA}{\mathrm{A}}

\newcommand{\LF}{\mathrm{F}}

\newcommand{\LR}{\mathrm{R}}
\newcommand{\LS}{\mathrm{S}}
\newcommand{\LT}{\mathrm{T}}
\newcommand{\LU}{\mathrm{U}}
\newcommand{\La}{\mathrm{a}}

\newcommand{\Lc}{\mathrm{c}}

\newcommand{\Ls}{\mathrm{s}}

\newcommand{\LCP}{{\textsc{lc}\raisebox{+0.5 pt}{$\scriptscriptstyle{+}$}}}

\newcommand{\cH}{\mathcal{H}}

\newcommand{\cU}{\mathcal{U}}
\newcommand{\cV}{\mathcal{V}}

\newcommand{\GeV}{\ \mathrm{GeV}}
\newcommand{\TeV}{\ \mathrm{TeV}}

\definecolor{red}{rgb}{1,0,0}
\newcommand{\NOTE}[1]{\textcolor{red}{ \bf[NOTE: #1]}}

\def\mi{{\mathrm i}}

\def\<>#1{\big\langle{#1}\big\rangle}
\def\[]#1{\big[{#1}\big]}

\newif\ifusefigs
\usefigstrue

\begin{document}

\title{Effect of color on rapidity gap survival}

\author{Zolt\'an Nagy}

\affiliation{
  DESY,
  Notkestrasse 85,
  22607 Hamburg, Germany
}

\email{Zoltan.Nagy@desy.de}

\author{Davison E. Soper}

\affiliation{
Institute of Theoretical Science,
University of Oregon,
Eugene, OR  97403-5203, USA
}

\email{soper@uoregon.edu}

\begin{abstract}
We study the probability for no jets with transverse momenta above a given cut to be found in the rapidity region between two high $P_\LT$ jets with a large rapidity separation. Our investigation uses the parton shower event generator \textsc{Deductor} with color beyond the leading-color-plus approximation included perturbatively.
\end{abstract}

\keywords{perturbative QCD, parton shower}
\preprint{DESY 19-082}
\date{8 August 2019}

\maketitle

\section{\label{sec:intro}Introduction}

In the collision of two high energy protons, it can happen that two partons scatter with a fairly small angle but still with substantial transverse momenta. This produces two high $P_\LT$ jets with a large difference in rapidity. At this Born level, there are no jets in the rapidity interval between the two high $P_\LT$ jets. We say that there is a rapidity gap. Further QCD radiation can produce jets in the gap region, so that the rapidity gap does not survive. 

In this paper, we investigate the role of color in the gap survival probability using the parton shower event generator \textsc{Deductor} \cite{NSI, NSII, NSspin, NScolor, Deductor, ShowerTime, PartonDistFctns, ColorEffects, NSThreshold, NSThresholdII}, which incorporates a systematically improvable approximation with respect to QCD color \cite{NSMoreColor}.

To state the physical problem precisely, consider events in proton collisions at $\sqrt s = 13 \TeV$. Using the anti-$k_\LT$ jet algorithm \cite{antikt} with a radius parameter $R$, find jets with transverse momenta $P_{\LT}$ and rapidities $y$ with $-Y_\mathrm{cut} < y < Y_\mathrm{cut}$. We will use $Y_\mathrm{cut} = 4.4$. Label the two highest $P_\LT$ jets 1 and 2, with $y_1 < y_2$. Define
\begin{equation}
\begin{split}
\bar p_\LT ={}& \frac{1}{2} (P_{\LT,1} + P_{\LT,2})
\;,
\\
y_{12} ={}& y_2 - y_1
\;.
\end{split}
\end{equation}
Now define a cut parameter $p_{\LT}^\mathrm{cut}$. We will take $p_{\LT}^\mathrm{cut} = 20 \GeV$. Look at those jets with $P_\LT >  p_{\LT}^\mathrm{cut}$ in the rapidity region $y_1 < y < y_2$ between the two leading jets. We will say that the event has a rapidity gap if there are no such jets in this rapidity region. 

For given values of $\bar p_\LT$ and $y_{12}$, let $f(\bar p_\LT, y_{12})$ be the fraction of events that have a gap. That is, $f$ is the ratio of the cross sections
\begin{equation}
f(\bar p_\LT, y_{12}) = 
\frac{d\sigma(\mathrm{gap})/[d\bar p_\LT\, dy_{12}]}
{d\sigma(\mathrm{total})/[d\bar p_\LT\, dy_{12}]}
\;.
\end{equation}
We can interpret $f$ as the probability that the gap survives after accounting for radiation beyond the Born level $2 \to 2$ scattering process. An alternative formulation is
\begin{equation}
\label{eq:fgapmod}
f(\bar p_\LT, y_{12}) = 
1 -
\frac{d\sigma(\mathrm{no\ gap})/[d\bar p_\LT\, dy_{12}]}
{d\sigma(\mathrm{total})/[d\bar p_\LT\, dy_{12}]}
\;.
\end{equation}
Here $d\sigma(\mathrm{no\ gap})/[d\bar p_\LT\, dy_{12}]$ is the cross section to have the two gap-defining jets plus at least one more jet with $P_\LT >  p_{\LT}^\mathrm{cut}$ in the gap region. This formulation is useful for perturbative calculations because both the numerator and the denominator in the second term are infrared safe cross sections that can be calculated at next-to-leading order (NLO).

There is a practical reason to explore the calculation of the gap fraction $f$.  In experimental investigations, it is often useful to look at some number of jets with transverse momenta $P_\LT \sim Q$, where $Q$ is large, say hundreds of GeV. These high $P_\LT$ jets can be a signal of physics beyond the Standard Model and are the objects of primary interest. In order to reduce backgrounds, it may be useful to impose a requirement that there be no jets beyond this that have  $P_\LT$ greater than some value $p_{\LT}^\mathrm{cut}$, where $p_{\LT}^\mathrm{cut}\ll Q$. If one does this, one needs to be able to estimate the fraction of signal events with no extra jets and the fraction of background events with no extra jets. The calculation of these fractions involves potential large logarithms, $\log(Q/p_{\LT}^\mathrm{cut})$. The large logarithms can spoil the usefulness of a calculation at a fixed order of perturbation theory. One can try to sum the large logarithms with an analytic calculation, but, as we suggest below, this is not entirely straightforward. An alternative is to use a parton shower event generator. This is the subject of this paper. The gap cross section defined above is the simplest example of a cross section that involves vetoing against extra jets.

There is also a motivation within QCD theory for examining the behavior of the gap fraction $f(\bar p_\LT, y_{12})$. In the case that $y_{12}$ is large, the behavior of $f$ as a function of $\bar p_\LT$ and $y_{12}$ is a matter of substantial theoretical interest because it brings together several issues concerning the structure of QCD. 

The perturbative expansion of the gap fraction $f$ contains two sorts of large logarithms. First, the logarithm $\log(\bar p_\LT/p_{\LT}^\mathrm{cut})$ can be large. Second, the rapidity separation $y_{12}$, which plays the role of a logarithm, can be large. At order $\as^N$, a perturbative calculation can give us a factor of $[y_{12} \times \log(\bar p_\LT/p_{\LT}^\mathrm{cut})]^N$, so a summation of large logarithms is called for. 

The summation of the large logarithms in $f$ is reviewed in Ref.~\cite{Manchester2009}. In the simplest approximation for an analytic summation of leading logarithms \cite{EarlyGap1, EarlyGap2}, one uses the exponential of a Sudakov exponent constructed from the one loop graphs for the virtual exchange of a low transverse momentum gluon. However, the analytic treatment is not straightforward \cite{Manchester2005}, so that a complete analytic summation is not available. The logarithms to be summed are ``non-global'' in that emissions into the gap region count differently from emissions outside of the gap region \cite{NonGlobal1, NonGlobal2, NonGlobal3, NonGlobal4}. There are imaginary contributions, in which a factor $y_{12}$ is replaced by a factor $\mi\pi$. Furthermore, in some contributions, a factor of $y_{12}$ or $\mi\pi$ becomes a factor of $\log(\bar p_\LT/p_{\LT}^\mathrm{cut})$ \cite{Manchester2009, SuperLeading1, SuperLeading2, Manchester2011}.

An alternative to an analytical summation of the large logarithms associated with $f$ is the use of a parton shower event generator like \textsc{Deductor}, which we use in this paper. We expect this to be useful because the splitting functions used in such a generator reflect the soft and collinear singularities that lead to the large logarithms. Furthermore, a parton shower treatment conserves momentum exactly at each step, whereas analytic treatments sometimes neglect the momentum of soft gluons.

One can worry that factors of $y_{12}$ that arise from integrating emissions over a range $y_1 < y < y_2$ may not be properly generated in a hardness ordered shower like \textsc{Deductor} if emitted gluons have roughly the same transverse momentum $P_\LT$. Perhaps it would be better to use a shower based on evolution in rapidity like \textsc{HEJ} \cite{HEJ, HEJgap}. However, with the ordering variable $\Lambda^2$ that is the default in \textsc{Deductor} and is used in this paper, it is possible to have successive gluon emissions with similar $P_\LT$ values if the rapidities $y$ of the emitted gluons are very different. This kinematic feature is discussed in some detail in Ref.~\cite{ShowerTime}.

The partons in the developing event radiate because they carry QCD color. In the very simplest approximation, the probability for gluon radiation from a quark is proportional to $C_\LF = (N_\Lc^2 - 1)/(2 N_\Lc)$ and the probability for gluon radiation from a gluon is proportional to $C_\LA = N_\Lc$, where $N_\Lc = 3$ is the number of colors. In a somewhat more sophisticated calculation, one can calculate emission probabilities in the leading color (LC) approximation, keeping contributions of leading order in an expansion in powers of $1/N_\Lc^2$. However, it is not self-evident that the leading color approximation is adequate for such a calculation. For instance, the perturbative expansion of the gap fraction can contain terms of the form $(1/N_\Lc^2)[\as y_{12} \log(\bar p_\LT/p_{\LT}^\mathrm{cut})]^N$. If we don't work beyond leading color, we lose such contributions.

In order to study the effect on the gap fraction of color beyond the leading color approximation, we use $\textsc{Deductor}$. The base color approximation in $\textsc{Deductor}$ is the LC+ approximation \cite{NScolor}, which includes some contributions that are suppressed by factors of $1/N_\Lc^2$. The effects of using the LC+ approximation are described in Ref.~\cite{ColorEffects}. With the current version\footnote{Version 3.0.3 of the code, 
  used in this paper, is available at 
  \href{http://www.desy.de/~znagy/deductor/}
  {http://www.desy.de/$\sim$znagy/deductor/}
  and
  \href{http://pages.uoregon.edu/soper/deductor/}
  {http://pages.uoregon.edu/soper/deductor/}.}  
of $\textsc{Deductor}$, we can go beyond the LC+ approximation. As explained in Ref.~\cite{NSMoreColor}, the operator that generates parton splittings with exactly the color content dictated by QCD Feynman diagrams is an operator denoted $\cH_I(t)$. In the LC+ approximation, $\cH_I(t)$ is approximated by an operator $\cH^\LCP(t)$ that has a simpler color structure. To get from the LC+ approximation to full color for splittings, we need another operator, $\Delta\cH(t)$, defined by
\begin{equation}
\label{eq:Hcolor}
\cH_I(t) = \cH^\LCP(t) 
+ \Delta\cH(t)
\;.
\end{equation}
Similarly, the operator that generates approximate virtual graphs in a shower with exactly the color content dictated by QCD Feynman diagrams is an operator denoted $\cV(t)$. This operator is used to construct the Sudakov factor for each shower step. In the LC+ approximation, $\cV(t)$ is approximated by an operator $\cV^\LCP(t)$. To get to full color for the virtual diagrams, we need another operator, $\Delta\cV(t)$, defined by
\begin{equation}
\label{eq:Vcolor}
\cV(t) = \cV^\LCP(t) 
+ \Delta\cV(t)
\;.
\end{equation}
The added contribution $\Delta\cV(t)$ includes an operator that contains a factor $\mi \pi$:
\begin{equation}
\label{eq:DelatVcolor}
\Delta\cV(t) = 
\Delta\cV_\mathrm{Re}(t)
+ \cV_{\mi\pi}(t)
\;.
\end{equation}

Now, \textsc{Deductor} allows one to include as many powers of $\Delta\cH(t)$ and $\Delta\cV(t)$ as one wants, within practical limits. In the calculations in this paper, we first investigate how many powers we need and then use just that number.

The commonly used parton shower algorithms \textsc{Pythia} \cite{Pythia}, \textsc{Herwig} \cite{Herwig}, and \textsc{Sherpa} \cite{Sherpa} work in the leading color approximation. There has been work other than ours on extending the accuracy of parton shower algorithms beyond the leading color approximation. Ref.~\cite{Seymour2018} works with color amplitudes, as in \cite{NSMoreColor}, but does not account for collinear singularities and so far lacks a parton shower implementation. Refs.~\cite{PlatzerSjodahl,  Isaacson:2018zdi, PlatzerSjodahlThoren} treat $\Delta \cH(t)$, but not $\Delta \cV(t)$.

This paper is structured as follows. There are three sections with preparatory information: Sec.~\ref{sec:bins} about putting the events in bins in $\bar p_\LT$, Sec.~\ref{sec:deductor} about \textsc{Deductor}, and Sec.~\ref{sec:colororder} about how many powers of $\Delta\cH(t)$ and $\Delta\cV(t)$ we need. Then Sec.~\ref{sec:gapfraction} contains the results on the gap fraction $f$ as a function of $\bar p_\LT$ and $y_{12}$. This includes results about the dependence of $f$ on the jet size parameter $R$, a comparison of the parton shower calculation to a purely perturbative calculation, and a comparison to results from \textsc{Pythia}. Finally, Sec.~\ref{sec:conclusions} presents some conclusions.

\section{\label{sec:bins}Putting the calculation in bins}

The gap fraction is a function $f(\bar p_\LT,y_{12})$ of the average transverse momentum of the jets that define the gap region and of their rapidity difference. It is defined by
\begin{equation}
f(\bar p_\LT,y_{12}) = \frac{d\sigma(\mathrm{gap})/[d\bar p_\LT\,dy_{12}]}
{d\sigma(\mathrm{total})/[d\bar p_\LT\,dy_{12}]}
\;.
\end{equation}
We organize the calculation of $f$ in bins of $\bar p_\LT$ and $y_{12}$: $ P_i < \bar p_\LT < P_{i+1}$, $Y_n < y_{12} < Y_{n+1}$. For each bin, the ratio that we calculate is
\begin{equation}
\begin{split}
f ={}&
\left[\int_{Y_n}^{Y_{n+1}}\!dy_{12}
\int_{P_i}^{P_{i+1}}\!d\bar p_\LT\
\frac{\sigma(\mathrm{gap})}{d\bar p_\LT\,dy_{12}}\,\frac{1}{h(n,\bar p_\LT)}
\right]
\\
&\bigg /
\left[\int_{Y_n}^{Y_{n+1}}\!dy_{12}
\int_{P_i}^{P_{i+1}}\!d\bar p_\LT\
\frac{\sigma(\mathrm{total})}{d\bar p_\LT\,dy_{12}}\,\frac{1}{h(n,\bar p_\LT)}
\right]
.
\end{split}
\end{equation}
Here the function $h(n,\bar p_\LT)$ is chosen so that $(1/h(n,\bar p_\LT))\,d\sigma(\mathrm{total})/[d\bar p_\LT\,dy_{12}]$ is approximately constant inside the bin. This gives us
\begin{equation}
\begin{split}
f ={}&
\int_{Y_n}^{Y_{n+1}}\!dy_{12}
\int_{P_i}^{P_{i+1}}\!d\bar p_\LT\
f(\bar p_\LT,y_{12})\,w(\bar p_\LT,y_{12})
\;,
\end{split}
\end{equation}
where the weight factor $w$ is
\begin{equation}
w(\bar p_\LT,y_{12}) = \frac{1}{N}
\frac{\sigma(\mathrm{total})}{d\bar p_\LT\,dy_{12}}\,
\frac{1}{h(n,\bar p_\LT)}
\;,
\end{equation}
with
\begin{equation}
N = \int_{Y_n}^{Y_{n+1}}\!dy_{12}
\int_{P_i}^{P_{i+1}}\!d\bar p_\LT\
\frac{\sigma(\mathrm{total})}{d\bar p_\LT\,dy_{12}}\,
\frac{1}{h(n,\bar p_\LT)}
\;,
\end{equation}
so that the integral of $w$ over the bin equals 1.

\section{\label{sec:deductor} The \textsc{Deductor} shower}

Our analysis is based on the parton shower event generator, \textsc{Deductor} \cite{NSI, NSII, NSspin, NScolor, Deductor, ShowerTime, PartonDistFctns, ColorEffects, NSThreshold, NSThresholdII, NSMoreColor}.  In this section, we review some of the features of \textsc{Deductor} that are particularly relevant to the gap survival problem.

The shower begins after a leading order $2 \to 2$ hard scattering. It would be desirable to use a next-to-leading order hard scattering with matching to the shower, but this option is not yet available in \textsc{Deductor}. For the hard scattering, we choose renormalization and factorization scales $\mu_R =  \mu_\LF = P_\LT^{\rm Born}/\sqrt{2}$, where $P_\LT^{\rm Born}$ is the transverse momentum in the Born scattering that initiates the shower.

We use the default shower ordering variable in \textsc{Deductor}, $\Lambda$, which is based on virtuality. For massless partons, the definition is
\begin{equation}
\begin{split}
\label{eq:Lambdadef}
\Lambda^2 ={}& \frac{(\hat p_l + \hat p_{m+1})^2}{2 p_l\cdot Q_0}\ Q_0^2
\hskip 1 cm {\rm final\ state},
\\
\Lambda^2 ={}&  \frac{|(\hat p_\La - \hat p_{m+1})^2|}{2 p_\La \cdot Q_0}\ Q_0^2
\hskip  1 cm {\rm initial\ state}.
\end{split}
\end{equation}
Here the mother parton in a final state splitting has momentum $p_l$ and the daughters have momenta $\hat p_l$ and $\hat p_{m+1}$. For an initial state splitting in hadron A, the mother parton has momentum $p_\La$, the new (in backward evolution) initial state parton has momentum $\hat p_\La$ and the final state parton created in the splitting has momentum $\hat p_{m+1}$. We denote by $Q_0$ a fixed vector equal to the total momentum of all of the final state partons just after the hard scattering that initiates the shower. The motivation for this choice is described in Ref.~\cite{ShowerTime}.

Successive splittings have $\Lambda_{n+1} < \Lambda_{n}$. For the first splitting, we demand that $\Lambda$ be smaller than a chosen shower start scale, $\mu_\Ls$. We choose 
\begin{equation}
\mu_\Ls = \frac{3}{2}\,P_\LT^{\rm Born}
\;.
\end{equation}
This choice is motivated in Ref.~\cite{NSThresholdII}. Results do not change much if we choose a larger value for $\mu_\Ls$ because the \textsc{Deductor} splitting kernel restricts the splitting transverse momentum to be no greater than $P_\LT^{\rm Born}$ in order to ensure that the scattering that initiates the shower is the highest $P_\LT$ scattering in the event.

The treatment of color in \textsc{Deductor} is described in detail in Ref.~\cite{NSMoreColor}. 

The base color treatment in \textsc{Deductor} is the LC+ approximation \cite{NScolor}. The LC+ approximation produces terms whose contributions are suppressed by powers $1/N_\Lc^n$, where $N_\Lc = 3$ is the number of colors. There is no need to carry terms suppressed by large powers of $1/N_\Lc$, so
we impose a maximum value on the color suppression index $I$ associated with a partonic color state \cite{NScolor}. Contributions to cross sections with a given value $I$ of the color suppression index come with a factor $1/N_\Lc^n$ with $n \ge I$. Thus we can neglect contributions with large values of $I$. We choose a value for a parameter $I_\mathrm{max}$. In this paper, we choose  $I_\mathrm{max} = 4$. The shower operator switches its behavior if it reaches a value of $I$ with $I - I_\mathrm{hard} \ge I_\mathrm{max}$, were $I_\mathrm{hard}$ is the color suppression index of the hard scattering state at the start of the shower. First, the shower switches to an approximate shower based on the color group $\LU(N_\Lc)$ instead of $\LS\LU(N_\Lc)$. Second, splittings that would increase $I$ are not allowed. Thus contributions proportional to $1/N_\Lc^{I_\mathrm{max}}$ are calculated only approximately.

The LC+ approximation is a substantial improvement over the leading color (LC) approximation, but it still leaves a lot out. What it leaves out are color operators $\Delta \cH$, $\Delta \cV_\mathrm{Re}$ and $\cV_{\mi\pi}$ from Eqs.~(\ref{eq:Hcolor}), (\ref{eq:Vcolor}) and (\ref{eq:DelatVcolor}) \cite{NSMoreColor}. A calculation that included all powers of these operators would be exact in color. \textsc{Deductor} cannot do that, but it can include a user-specified maximum number of powers of $\Delta \cH$, $\Delta \cV_\mathrm{Re}$ and $\cV_{\mi\pi}$. Here $\Delta \cH$ is the part of parton splitting graphs that is omitted in the LC+ approximation; $\Delta \cV_\mathrm{Re}$ is the real part of approximated virtual graphs omitted in the LC+ approximation; and $\cV_{\mi\pi}$ is the imaginary part of virtual graphs, which contain a factor $\mi \pi$.

The user controls the level of approximation by specifying integers $N_\Delta^\mathrm{thr}$, $N_\mathrm{Re}$, and $N_{\mi\pi}$. 

First, just after the hard scattering, \textsc{Deductor} inserts an operator $\cU_\cV$ that produces a summation of threshold logarithms \cite{NSThresholdII}. The operator $\cU_\cV$ gives results as an expansion in powers of $\Delta \cV_\mathrm{Re}$. \textsc{Deductor} retains only those terms with no more than $N_\Delta^\mathrm{thr}$ factors of $\Delta \cV_\mathrm{Re}$. 

Second, the operators $\Delta \cH$, $\Delta \cV_\mathrm{Re}$ and $\cV_{\mi\pi}$ appear in the shower evolution operator $\cU(t_2,t_1)$. The operator $\cU(t_2,t_1)$ produces terms proportional to $[\Delta \cH]^A [\Delta \cV_\mathrm{Re}]^B [\cV_{\mi \pi}]^C$. \textsc{Deductor} retains only terms with $A+B \le N_\mathrm{Re}$, $C \le N_{\mi\pi}$, and $A + B + C \le \max\{N_\mathrm{Re},N_{\mi\pi}\}$.

Finally, if the shower evolution reaches a state in which the color suppression index $I$ has $I - I_\mathrm{hard} \ge I_\mathrm{max}$, then the evolution omits any further contributions from $\Delta \cH$, $\Delta \cV_\mathrm{Re}$ and $\cV_{\mi\pi}$ and switches to a $\LU(N_\Lc)$ instead of an $\LS\LU(N_\Lc)$ shower, while not allowing $I$ to increase further.

\textsc{Deductor} allows the user to specify $N_\Delta^\mathrm{thr}$ at the start of the parton shower. Then the user can specify approximation parameters $N_\mathrm{Re}$, $N_{\mi\pi}$ and $I_\mathrm{max}$ for an evolution interval $\mu_\Ls > \Lambda > \Lambda(1)$ and then smaller parameters for successive following evolution intervals $\Lambda(i) > \Lambda > \Lambda(i+1)$. In this paper, we use parameters $N_\mathrm{Re}$, $N_{\mi\pi}$ and $I_\mathrm{max} = 4$ for the interval $\mu_\Ls > \Lambda > \Lambda(1) = 30 \GeV$. Then we either stop the shower at 30 GeV or continue it to $\Lambda(2) = 1 \GeV$ with the LC+ approximation, $N_\mathrm{Re} = N_{\mi\pi} = 0$, with, still, $I_\mathrm{max} = 4$.

The \textsc{Deductor} splitting kernel also limits the transverse momenta $k_\LT$ in splittings to be larger than $k_\LT^{\rm min} \approx 1 \GeV$. The precise value is  $k_\LT^{\rm min} = 1.0 \GeV$ for final state splittings and $k_\LT^{\rm min} = 1.295 \GeV$, set by the starting scale of the parton distributions that we use, for initial state splittings.

\section{\label{sec:colororder}Effect of order of approximation}

As outlined above, we apply limits on the accuracy of the parton shower calculation by specifying parameters $N_\Delta^\mathrm{thr}$, $N_\mathrm{Re}$, $N_{\mi\pi}$ and $I_\mathrm{max}$, along with the range of the hardness variable $\Lambda$ over which the parameters apply. In this section, we try to estimate the systematic error in the calculation that results from these limits. 

For the color suppression index, we set $I_\mathrm{max} = 4$. This means that contributions to $f$ that carry a factor $1/N_\Lc^4 \approx 10^{-2}$ and beyond are calculated only approximately. Thus we estimate a systematic error on $f$ of $\pm 0.01$ from not taking a larger value of $I_\mathrm{max}$.

For our studies of the gap fraction, in the evolution range $\mu_\Ls > \Lambda > 30 \GeV$, we will choose $N_\mathrm{Re} = 2$, $N_{\mi\pi} = 2$, and $N_\Delta^\mathrm{thr} = 1$. How much systematic error should one ascribe to not choosing larger values to these parameters? We investigate that question in this section by changing these parameters one at a time, while leaving the remaining parameters set to zero. In this investigation, we examine $f$ in the range $300 \GeV < \bar p_\LT < 400 \GeV$ and $4 < y_{12} < 5$. We set $N_\mathrm{Re}$ and $N_{\mi\pi}$ in the evolution range $\mu_\Ls > \Lambda > 30 \GeV$ and stop evolution at $\Lambda = 30 \GeV$.

\begin{figure}
\begin{center}
\ifusefigs 

\begin{tikzpicture}
  \begin{axis}[title = {Effect of $N_\mathrm{Re}$},
    xlabel={$N_\mathrm{Re}$}, ylabel={$f$},
    xticklabels={0,1,2,3},
    xtick={0,1,2,3},
    ymin=0.10, ymax=0.22,
    ytick={0.10,0.12,0.14,0.16,0.18,0.20,0.22},
    yticklabels={0.10,0.12,0.14,0.16,0.18,0.20,0.22},
    ]
    
\addplot +[only marks, error bars/.cd, y dir = both, y explicit]
coordinates{
(0,0.199418)  +- (0,0.0012444)   
(1,0.158211)  +- (0,0.00398641)  
(2,0.18647)   +- (0,0.00900461)  
(3,0.191794)  +- (0,0.0203162)   
};

\end{axis}
\end{tikzpicture}

\else 
\NOTE{Figure fig:Nre goes here.}
\fi

\end{center}
\caption{
Gap fraction $f$ in the range $300 \GeV < \bar p_\LT < 400 \GeV$ and $4 < y_{12} < 5$ calculated with different values of $N_\mathrm{Re}$ with $N_{\mi\pi} = N_\Delta^\mathrm{thr} = 0$.
}
\label{fig:Nre}
\end{figure}
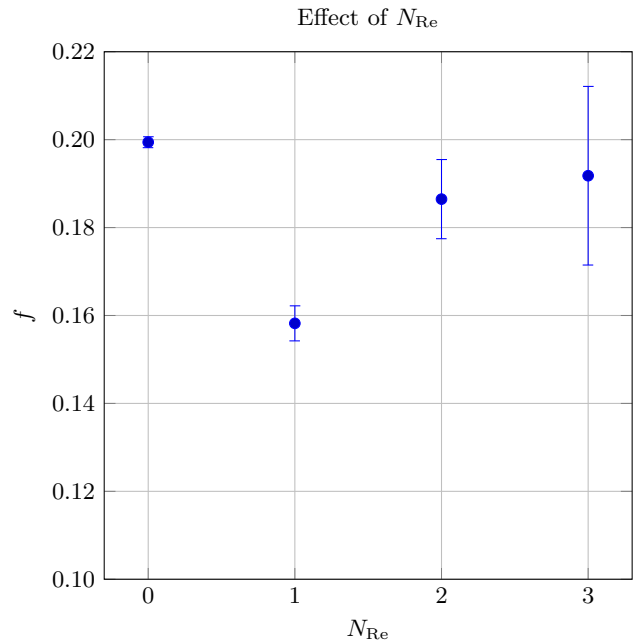

We first  investigate how $f$ depends on $N_\mathrm{Re}$, with $N_\Delta^\mathrm{thr} = N_{\mi\pi} = 0$. The results are shown in Fig.~\ref{fig:Nre}. The value of $f$ for $N_\mathrm{Re} = 0$ is the result of a calculation in the LC+ approximation. We see that adding one factor of $\Delta \cH$ or $\Delta \cV_\mathrm{Re}$ changes $f$ by $- 0.041 \pm 0.004$, approximately a 20\% change. Choosing $N_\mathrm{Re} = 2$ leads to a further change of $+ 0.028 \pm 0.010$.  We expect that increasing $N_\mathrm{Re}$ beyond 2 will change $f$ by less than 0.02. Indeed, we find that choosing $N_\mathrm{Re} = 3$ leads to a further change of $0.01 \pm 0.02$. Based on this study, we estimate that the error on $f$ from limiting $N_\mathrm{Re}$ to 2 is $\pm 0.02$.

\begin{figure}
\begin{center}
\ifusefigs 

\begin{tikzpicture}
  \begin{axis}[title = {Effect of $N_{\mi\pi}$},
    xlabel={$N_{\mi\pi}$}, ylabel={$f$},
    ymin=0.16, ymax=0.24,
    ytick={0.16,0.18,0.20,0.22,0.24},
    yticklabels={0.16,0.18,0.20,0.22,0.24},
    ]
    
\addplot +[only marks, error bars/.cd, y dir = both, y explicit]
coordinates{
(0,0.199418)  +- (0,0.00129594) 
(2,0.214931)  +- (0,0.00203912) 
(4,0.21485)   +- (0,0.00199111) 
};

\end{axis}
\end{tikzpicture}

\else 
\NOTE{Figure fig:NiPi goes here.}
\fi

\end{center}
\caption{
Gap fraction $f$ as in Fig.~\ref{fig:Nre} calculated with different values of $N_{\mi\pi}$ with $N_\mathrm{Re} = N_\Delta^\mathrm{thr} = 0$. 
}
\label{fig:NiPi}
\end{figure}
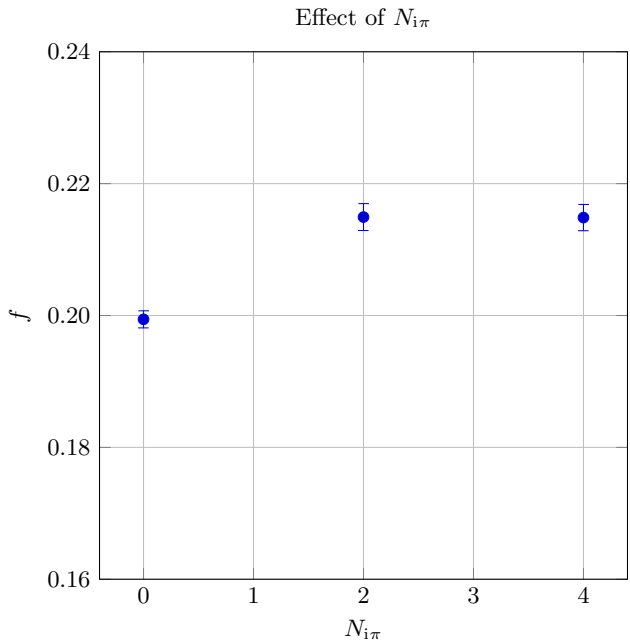

We next investigate how $f$ depends on $N_{\mi\pi}$ with $N_\mathrm{Re} = N_\Delta^\mathrm{thr} = 0$. The results are shown in Fig.~\ref{fig:NiPi}. Only even powers of $\cV_{\mi\pi}$ contribute to the cross section since $\cV_{\mi\pi}$ has a factor $\mi$. Thus we show results only for $N_{\mi\pi} = 0$, 2, and 4. We see that increasing $N_{\mi\pi}$ from 0 to 2 in $\cU$ raises the gap fraction by $0.016 \pm 0.002$. Adding two more powers of $\cV_{\mi\pi}$ leaves the gap fraction unchanged, within the statistical error. The change is $0.000 \pm 0.003$. We conclude that $N_{\mi\pi} = 2$ is a reasonable choice and that $\pm 0.01$ is a reasonable error estimate for the effect of added factors of $\cV_{\mi\pi}$ beyond 2.

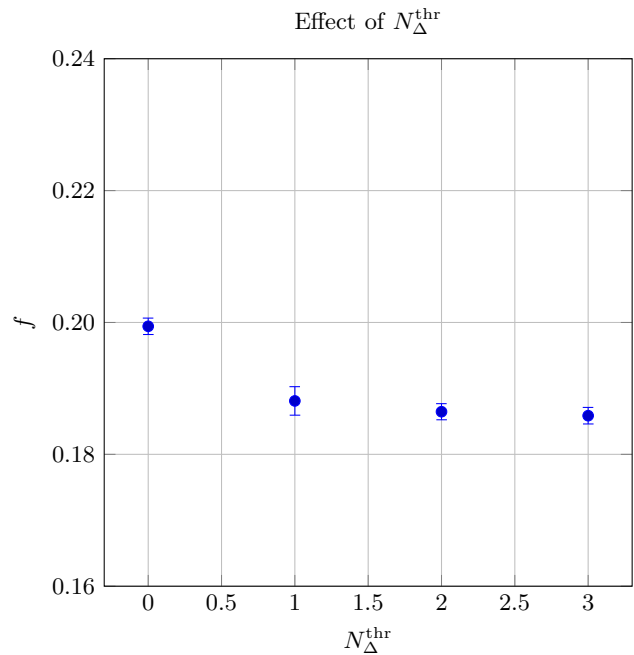
\begin{figure}
\begin{center}
\ifusefigs 

\begin{tikzpicture}
  \begin{axis}[title = {Effect of $N_\Delta^\mathrm{thr}$},
    xlabel={$N_\Delta^\mathrm{thr}$}, ylabel={$f$},
    ymin=0.16, ymax=0.24,
    ytick={0.16,0.18,0.20,0.22,0.24},
    yticklabels={0.16,0.18,0.20,0.22,0.24},
    ]
    
\addplot +[only marks, error bars/.cd, y dir = both, y explicit]
coordinates{
(0,0.199418)  +- (0,0.0012444)   
(1,0.188101)  +- (0,0.00216162)  
(2,0.186464)  +- (0,0.00122247)  
(3,0.185862)  +- (0,0.00125425)  
};

\end{axis}
\end{tikzpicture}

\else 
\NOTE{Figure fig:Nthr goes here.}
\fi

\end{center}
\caption{
Gap fraction $f$ as in Fig.~\ref{fig:Nre} calculated with different values of $N_\Delta^\mathrm{thr}$ with $N_\mathrm{Re} = N_{\mi\pi} = 0$.
}
\label{fig:Nthr}
\end{figure}

We next investigate how $f$ depends on $N_\Delta^\mathrm{thr}$ with $N_\mathrm{Re} = N_{\mi\pi} = 0$. The results are shown in Fig.~\ref{fig:Nthr}. We see that increasing $N_\Delta^\mathrm{thr}$ from 0 to 1 lowers the gap fraction by $0.011 \pm 0.002$. Increasing $N_\Delta^\mathrm{thr}$ above 1 has hardly any effect. We conclude that $N_\Delta^\mathrm{thr} = 1$ is a reasonable choice and that $\pm 0.01$ is a reasonable error estimate for the effect of added color in the threshold operator in this calculation. 

\begin{figure}
\begin{center}
\ifusefigs 

\begin{tikzpicture}
  \begin{axis}[title = {Effect of $\Lambda_\mathrm{min}$},
    xlabel={$\Lambda_\mathrm{min}$ (GeV)}, ylabel={$f$},
    xmin=0, xmax=50,
    ymin=0.1, ymax=0.25,
    ytick={0.10,0.15,0.20,0.25},
    yticklabels={0.10,0.15,0.20,0.25},
    ]
    
\addplot +[only marks, error bars/.cd, y dir = both, y explicit]
coordinates{
(20, 0.187209) +- (0,0.0232346) 
(30, 0.169261) +- (0,0.0201618) 
(40, 0.187136) +- (0,0.012831)  
};
    
\addplot[red, domain=1:49] 
{0.173217 + 0.000284714*x};

\end{axis}
\end{tikzpicture}

\else 
\NOTE{Figure fig:LambdaMin goes here.}
\fi

\end{center}
\caption{
Gap fraction $f$ in the range $300 \GeV < \bar p_\LT < 400 \GeV$ and $4 < y_{12} < 5$ as a function of $\Lambda_\mathrm{min}$. The calculation uses $N_\Delta^\mathrm{thr} = 1$ at the start of the shower, then $N_\mathrm{Re} = 2$, $N_{\mi\pi} = 2$ for the shower in the range $\mu_\Ls > \Lambda > \Lambda_\mathrm{min}$, then $N_\mathrm{Re} = N_{\mi\pi} = 0$ in the range $\Lambda_\mathrm{min} > \Lambda > 1 \GeV$. The maximum color suppression index is $I_\mathrm{max} = 4$. The curve is a linear fit to the numerical results.
}
\label{fig:LambdaMin}
\end{figure}
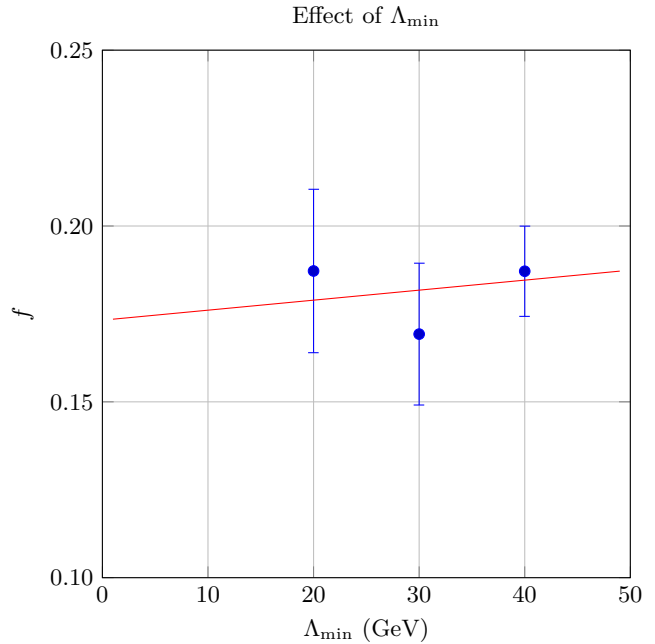

In the calculations of the gap fraction in the following section, we use $N_\mathrm{Re} = 2$, $N_{\mi\pi} = 2$, $N_\Delta^\mathrm{thr} = 1$ in the range $\mu_\Ls > \Lambda > \Lambda_\mathrm{min}$, where we choose $\Lambda_\mathrm{min} = 30 \GeV$. Then we revert to the LC+ approximation for $\Lambda_\mathrm{min} > \Lambda > 1 \GeV$. The range $\mu_\Ls > \Lambda > 30 \GeV$ covers a range of about a factor of 10 or more in $\Lambda$ for $\bar p_\LT > 200 \GeV$ (with $\mu_\Ls = 3 \bar p_\LT/2$). We hope that this range is adequate to explore the effect of extra color contributions. However, we will find that including color beyond the LC+ approximation generally makes $f$ smaller. Thus if we were able to include the extra color contributions in the range $30 \GeV > \Lambda > 1 \GeV$, presumably the calculated $f$ would be somewhat smaller. If we were to simply set $\Lambda_\mathrm{min} = 1 \GeV$, \textsc{Deductor} would generate very complicated color states in a large number of quite soft splittings, so that few events would result and the statistical fluctuations in $f$ would be large. An alternative is to try to estimate the effect of leaving $\Lambda_\mathrm{min} = 30 \GeV$ by extrapolating in $\Lambda_\mathrm{min}$. We calculate $f$ in the range $300 \GeV < \bar p_\LT < 400 \GeV$ and $4 < \Delta y < 5$ for $\Lambda_\mathrm{min} =$ 20 GeV, 30 GeV and 40 GeV. If we assume a linear model, $f = a_0 + a_1 \Lambda_\mathrm{min}/(30 \GeV)$, we can fit $a_0$ and $a_1$.  Then results for $f$ in this $\bar p_\LT$ and $y_{12}$ range calculated with $\Lambda_\mathrm{min} = 30 \GeV$ should be corrected by subtracting $a_1$ from $f$. The results for $f$ and the fit are shown in Fig.~\ref{fig:LambdaMin}. We find $a_1 = 0.01 \pm 0.04$. That is, $a_1$ equals 0 within its statistical error. However, with the accuracy obtained for $a_1$, we have an extrapolation error on $f$ in this $\bar p_\LT$ and $y_{12}$ range of about $\pm 0.04$.

\section{\label{sec:gapfraction} Results for the gap fraction}

We are now ready to look at the results for the gap fraction $f$. The jets are defined with the anti-$k_\LT$ algorithm. We start with radius parameter $R = 0.4$. Then we will examine how $f$ depends on $R$. We choose five different bins for $y_{12}$ and examine $f$ in each bin as a function of $\bar p_\LT$. 

We use $N_\mathrm{Re} = 2$, $N_{\mi\pi} = 2$, $N_\Delta^\mathrm{thr}=1$ and $I_\mathrm{max} = 4$. There are systematic errors in the results that arise from not using larger values of $N_\mathrm{Re}$, $N_{\mi\pi}$, $N_\Delta^\mathrm{thr}$ and $I_\mathrm{max}$. In Sec.~\ref{sec:colororder}, we estimated these systematic errors in $f$ at about $\pm 0.02$. 

In each case, the results were obtained with the stated values of $N_\mathrm{Re}$ and $N_{\mi\pi}$ in the shower between $\Lambda = \mu_\Ls$ and $\Lambda = \Lambda_\mathrm{min} = 30 \GeV$. For the rest of the shower, down to $\Lambda = 1 \GeV$, we used the LC+ approximation. There is a systematic error from not using a smaller value of $\Lambda_\mathrm{min}$, based on how well we could extrapolate to $\Lambda_\mathrm{min} = 1 \GeV$. In Sec.~\ref{sec:colororder}, we estimate this extrapolation error at $\pm 0.04$.

There are also systematic errors from not having shower splitting functions beyond order $\as$ and from starting the shower with just lowest order parton scattering. We do not estimate these systematic errors and, rather, regard the results as an investigation of color effects within a calculation at this order of approximation.

Finally, there are statistical errors from the fluctuations in Monte Carlo event generation. The statistical errors are rather substantial for the largest values of $y_{12}$ and $\bar p_\LT$. We do not exhibit error bands that represent the statistical errors since the size of the fluctuations is evident in the differences of $f$ between neighboring values of $\bar p_\LT$. \textsc{Deductor}, of course, provides an estimated error for each bin, but our impression is that these estimated errors are somewhat smaller than the bin to bin fluctuations.

\begin{figure}
\begin{center}
\ifusefigs 

\begin{tikzpicture}
 \begin{groupplot}[
      group style={
          group size=1 by 5,
          vertical sep=0pt,
          x descriptions at=edge bottom},
          xlabel={$\bar p_\LT\,\mathrm{[GeV]}$},
          width=8.6cm,
    ]


\nextgroupplot[
ylabel={$f(\bar p_\LT)$},
height=5cm,
xmin=50, xmax=500,
ymin=0.1, ymax=1.0,
xmode=log,
ymode=log,
ytick={0.1,0.2,0.3,0.4,0.5,0.6,0.7,0.8,0.9,1.0},
yticklabels={,0.2,0.3,0.4,,0.6,,0.8,,1.0},
xtick={50,60,70,80,90,100,200,300,400,500},
  legend cell align=left,
  every axis legend/.append style={
  at={(0.03,0.05)},
  anchor = south west},
]

\pgfplotstableread{
55.     0.840518
65.     0.798913
75.     0.763499
85.     0.734
97.5    0.69579
112.5   0.661739
127.5   0.638247
142.5   0.616068
165.    0.586371
195.    0.5526
225.    0.526907
255.    0.498009
285.    0.487869
320.    0.46635
360.    0.453832
400.    0.437371
440.    0.425733
480.    0.417114
}\ductLCpA

\addlegendentry{LC+}

\pgfplotstableread{
55.     0.842406
65.     0.799189
75.     0.76163
85.     0.73333
97.5    0.695443
112.5   0.65584
127.5   0.63153
142.5   0.618572
165.    0.584554
195.    0.537051
225.    0.525483
255.    0.490918
285.    0.505572
320.    0.471853
360.    0.448453
400.    0.437108
440.    0.414043
480.    0.42033
}\ductRealA

\addlegendentry{$\Delta\cH$, $\Delta \cV_\mathrm{Re}$}

\pgfplotstableread{
55.     0.84419
65.     0.804257
75.     0.768508
85.     0.740681
97.5    0.703389
112.5   0.667508
127.5   0.634201
142.5   0.634772
165.    0.603499
195.    0.552161
225.    0.545113
255.    0.508077
285.    0.519732
320.    0.48743
360.    0.463172
400.    0.453249
440.    0.427985
480.    0.436254
}\ductAllA

\addlegendentry{$\Delta\cH$, $\Delta \cV_\mathrm{Re}$,$\cV_{\mi\pi}$ }

\addplot [blue, semithick] table [x={0},y={1}]{\ductLCpA};

\addplot [darkgreen, semithick] table [x={0},y={1}]{\ductRealA};

\addplot [red, semithick] table [x={0},y={1}]{\ductAllA};

\node at (350.0,0.8) {\fcolorbox{black}{white}{\small{$1 < y_{12} < 2$}}};


\nextgroupplot[
ylabel = $f(\bar p_\LT)$,
height = 5 cm,
xmin=50, xmax=500,
ymin=0.1, ymax=1.0,
xmode=log,
ymode=log,
ytick={0.1,0.2,0.3,0.4,0.5,0.6,0.7,0.8,0.9,1.0},
yticklabels={,0.2,0.3,0.4,,0.6,,0.8,,1.0},
xtick={50,60,70,80,90,100,200,300,400,500},
]

\pgfplotstableread{
55.     0.764878
65.     0.711129
75.     0.659845
85.     0.63029
97.5    0.577288
112.5   0.537826
127.5   0.501771
142.5   0.479551
165.    0.441214
195.    0.411196
225.    0.38228
255.    0.360394
285.    0.343351
320.    0.329854
360.    0.314727
400.    0.297306
440.    0.290362
480.    0.283321
}\ductLCpB

\pgfplotstableread{
55.     0.769368
65.     0.708563
75.     0.654098
85.     0.630535
97.5    0.57621
112.5   0.528456
127.5   0.494705
142.5   0.475869
165.    0.434734
195.    0.401549
225.    0.367705
255.    0.354799
285.    0.340477
320.    0.307931
360.    0.312532
400.    0.27857
440.    0.265303
480.    0.244131
}\ductRealB

\pgfplotstableread{
55.     0.775029
65.     0.716019
75.     0.66219
85.     0.642467
97.5    0.590053
112.5   0.542435
127.5   0.514784
142.5   0.494494
165.    0.453119
195.    0.421344
225.    0.387949
255.    0.37101
285.    0.359428
320.    0.326792
360.    0.331543
400.    0.29846
440.    0.286284
480.    0.26211
}\ductAllB

\addplot [blue, semithick] table [x={0},y={1}]{\ductLCpB};

\addplot [darkgreen, semithick] table [x={0},y={1}]{\ductRealB};

\addplot [red, semithick] table [x={0},y={1}]{\ductAllB};

\node at (350.0,0.8) {\fcolorbox{black}{white}{\small{$2 < y_{12} < 3$}}};


\nextgroupplot[
ylabel = $f(\bar p_\LT)$,
height = 5 cm,
xmin=50, xmax=500,
ymin=0.1, ymax=1.0,
xmode=log,
ymode=log,
ytick={0.1,0.2,0.3,0.4,0.5,0.6,0.7,0.8,0.9,1.0},
yticklabels={,0.2,0.3,0.4,,0.6,,0.8,,1.0},
xtick={50,60,70,80,90,100,200,300,400,500},]

\pgfplotstableread{
55.     0.696003
65.     0.620445
75.     0.570868
85.     0.53878
97.5    0.491351
112.5   0.450772
127.5   0.42062
142.5   0.378167
165.    0.357102
195.    0.321557
225.    0.291665
255.    0.276664
285.    0.264924
320.    0.244382
360.    0.234631
400.    0.227397
440.    0.219717
480.    0.217836
}\ductLCpC

\pgfplotstableread{
55.     0.698916
65.     0.628799
75.     0.566458
85.     0.542248
97.5    0.495019
112.5   0.438461
127.5   0.389986
142.5   0.384994
165.    0.352162
195.    0.304028
225.    0.268094
255.    0.260305
285.    0.250634
320.    0.227774
360.    0.222138
400.    0.201605
440.    0.200021
480.    0.202425
}\ductRealC

\pgfplotstableread{
55.     0.705436
65.     0.640328
75.     0.583098
85.     0.555647
97.5    0.511811
112.5   0.458993
127.5   0.405698
142.5   0.404044
165.    0.369061
195.    0.318085
225.    0.290369
255.    0.276893
285.    0.26666
320.    0.245575
360.    0.237632
400.    0.220866
440.    0.218623
480.    0.222181
}\ductAllC

\addplot [blue, semithick] table [x={0},y={1}]{\ductLCpC};

\addplot [darkgreen, semithick] table [x={0},y={1}]{\ductRealC};

\addplot [red, semithick] table [x={0},y={1}]{\ductAllC};

\node at (350.0,0.8) {\fcolorbox{black}{white}{\small{$3 < y_{12} < 4$}}};


\nextgroupplot[
ylabel = $f(\bar p_\LT)$,
height = 5 cm,
xmin=50, xmax=500,
ymin=0.1, ymax=1.0,
xmode=log,
ymode=log,
ytick={0.1,0.2,0.3,0.4,0.5,0.6,0.7,0.8,0.9,1.0},
yticklabels={,0.2,0.3,0.4,,0.6,,0.8,,1.0},
xtick={50,60,70,80,90,100,200,300,400,500},]

\pgfplotstableread{
55.     0.614134
65.     0.553826
75.     0.517392
85.     0.479122
97.5    0.425718
112.5   0.401904
127.5   0.357247
142.5   0.327625
165.    0.299872
195.    0.275256
225.    0.253303
255.    0.23927
285.    0.224482
320.    0.217796
360.    0.212131
400.    0.207431
440.    0.204041
480.    0.207425
}\ductLCpD

\pgfplotstableread{
55.     0.630287
65.     0.567355
75.     0.543419
85.     0.48177
97.5    0.431408
112.5   0.412216
127.5   0.365044
142.5   0.332189
165.    0.289021
195.    0.254083
225.    0.21641
255.    0.189779
285.    0.196343
320.    0.177018
360.    0.185339
400.    0.143766
440.    0.145101
480.    0.171019
}\ductRealD

\pgfplotstableread{
55.     0.638322
65.     0.579003
75.     0.552482
85.     0.494077
97.5    0.442697
112.5   0.423841
127.5   0.379623
142.5   0.349576
165.    0.299791
195.    0.269318
225.    0.23228
255.    0.203213
285.    0.214877
320.    0.192609
360.    0.204034
400.    0.160387
440.    0.165567
480.    0.193562
}\ductAllD

\addplot [blue, semithick] table [x={0},y={1}]{\ductLCpD};

\addplot [darkgreen, semithick] table [x={0},y={1}]{\ductRealD};

\addplot [red, semithick] table [x={0},y={1}]{\ductAllD};

\node at (350.0,0.8) {\fcolorbox{black}{white}{\small{$4 < y_{12} < 5$}}};


\nextgroupplot[
ylabel={$f(\bar p_\LT)$},
height=5cm,
xmin=50, xmax=500,
ymin=0.1, ymax=1.0,
xmode=log,
ymode=log,
ytick={0.1,0.2,0.3,0.4,0.5,0.6,0.7,0.8,0.9,1.0},
yticklabels={0.1,0.2,0.3,0.4,,0.6,,0.8,,1.0},
xtick={50,60,70,80,90,100,200,300,400,500},
xticklabels={50,60,,80,,100,200,300,400,500},]

\pgfplotstableread{
55.     0.534401
65.     0.493032
75.     0.449917
85.     0.434367
97.5    0.391802
112.5   0.360165
127.5   0.337951
142.5   0.307465
165.    0.291556
195.    0.269378
225.    0.246162
255.    0.255607
285.    0.237343
320.    0.2448
360.    0.254609
400.    0.246019
440.    0.256981
480.    0.282283
}\ductLCpE

\pgfplotstableread{
55.     0.546572
65.     0.489224
75.     0.445746
85.     0.421993
97.5    0.419914
112.5   0.372338
127.5   0.33981
142.5   0.250403
165.    0.261904
195.    0.241868
225.    0.20227
255.    0.192121
285.    0.190201
320.    0.15512
360.    0.185308
400.    0.151707
440.    0.131917
480.    0.168073
}\ductRealE

\pgfplotstableread{
55.     0.552585
65.     0.493972
75.     0.452348
85.     0.424831
97.5    0.422773
112.5   0.367604
127.5   0.349166
142.5   0.25376
165.    0.271546
195.    0.253071
225.    0.209343
255.    0.210271
285.    0.199244
320.    0.172916
360.    0.202647
400.    0.174901
440.    0.149248
480.    0.194381
}\ductAllE

\addplot [blue, semithick] table [x={0},y={1}]{\ductLCpE};

\addplot [darkgreen, semithick] table [x={0},y={1}]{\ductRealE};

\addplot [red, semithick] table [x={0},y={1}]{\ductAllE};

\node at (350.0,0.8) {\fcolorbox{black}{white}{\small{$5 < y_{12} < 6$}}};
\end{groupplot}
\end{tikzpicture}

\else 
\NOTE{Figure fig:fR04 goes here.}
\fi

\end{center}
\caption{
Gap fraction $f$ for $R = 0.4$ versus $y_{12}$ and $\bar p_\LT$. 
}
\label{fig:fR04}
\end{figure}
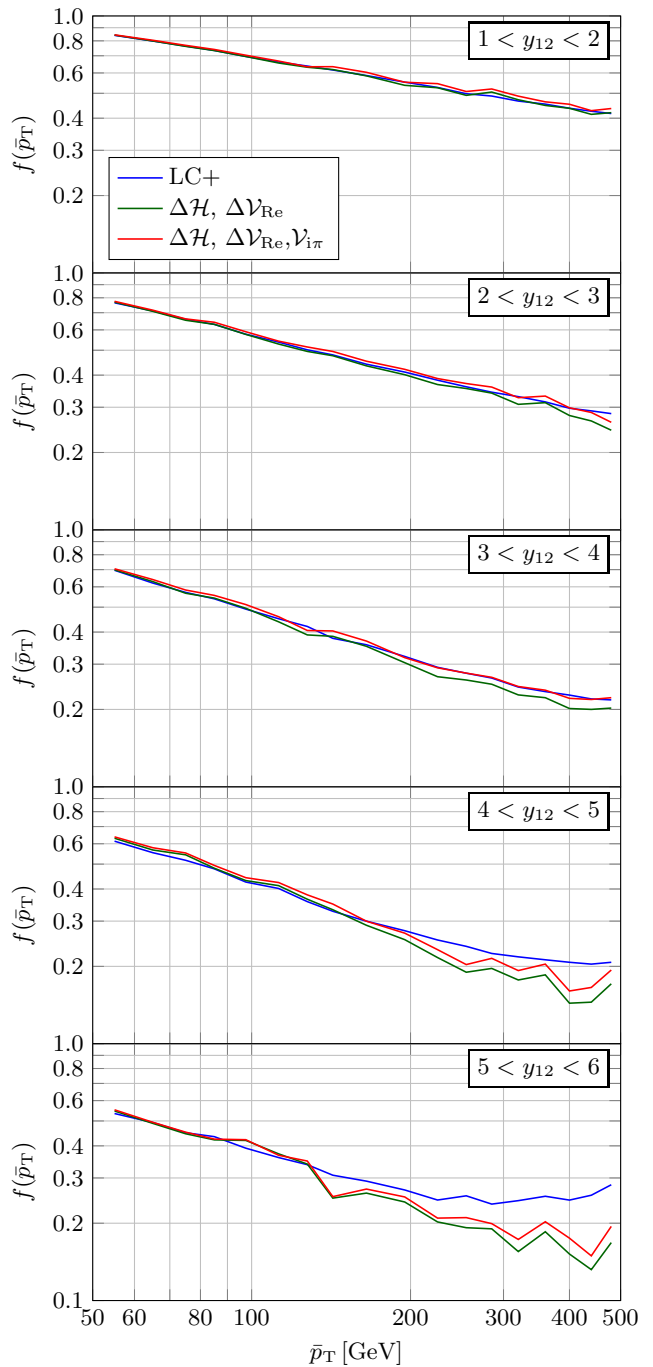

We now turn to the results. 

\subsection{\label{sec:R04}Gap fraction for $R = 0.4$}
\label{sec:gap04}

 We begin with results for jets defined with the anti-$k_\LT$ algorithm with a radius parameter $R = 0.4$, shown in Fig.~\ref{fig:fR04}. We choose five different bins for $y_{12}$ and examine $f(\bar p_\LT)$ as a function of $\bar p_\LT$. In each $y_{12}$ bin, we show three curves, all obtained with $I_\mathrm{max} = 4$. The first, in blue, is obtained with just the LC+ approximation. Then, in green, we show results obtained with contributions from $\Delta \cH$ and $\Delta \cV_\mathrm{Re}$ using $N_\mathrm{Re} = 2$ and $N_\Delta^\mathrm{thr}=1$. Here contributions from $\cV_{\mi \pi}$ are omitted. Finally, in red, we show results obtained with contributions from all of $\Delta \cH$, $\Delta \cV_\mathrm{Re}$ and $\cV_{\mi\pi}$ using $N_\mathrm{Re} = 2$, $N_{\mi\pi} = 2$, and $N_\Delta^\mathrm{thr}=1$.

Look first at $f(\bar p_\LT)$ for $1 < y_{12} < 2$. We see that all three plots are almost straight lines. That is, $f$ has the approximate form
\begin{equation}
\label{eq:fexponential}
f(\bar p_\LT) \approx \exp(- A_0 - A_1 \log(\bar p_\LT/p_\LT^\mathrm{cut}))
\;.
\end{equation}
We find that $A_1 \approx 0.33$. It appears, then, that the dependence of $f$ on the large logarithm $\log(\bar p_\LT/p_\LT^\mathrm{cut})$ exponentiates. We also note that the three curves are almost identical: the contributions from the operators $\Delta \cH$, $\Delta \cV_\mathrm{Re}$ and $\cV_{\mi\pi}$ are quite small.

Look next at $f(\bar p_\LT)$ for $2 < y_{12} < 3$. We see that all three plots are almost straight lines. The slope of the lines is larger: for LC+, $A_1 \approx 0.50$. This is what we expect. There is now a larger gap to radiate gluons into, so it is easier for the gap not to survive. 

We also see that the contributions from $\Delta\cH$ and $\Delta\cV_\mathrm{Re}$ have begun to matter. Including the effects of these operators multiplies $f$ by a factor of about $\exp(- 0.13)$ in the region $\bar p_\LT > 300 \GeV$. Some effect like this was to be expected. The operator $\Delta\cH$ gives soft gluon emission in the angular region between two emitting partons, one in the ket state and one in the bra state. Such a gluon can destroy the gap. There is now a wider angular region for this to happen, so it is not unexpected that an effect of $\Delta\cH$ and $\Delta\cV_\mathrm{Re}$ would begin to be visible.

The contribution from $\cV_{\mi \pi}^2$ is visible as the difference between the red and green curves in Fig.~\ref{fig:fR04}. We see that for $2 < y_{12} < 3$, the contribution from $\cV_{\mi \pi}^2$ has begun to matter. The operator $\cV_{\mi \pi}$ does not create any final state gluons. However, it changes the color state of the color density matrix. With different color, the probability for the other operators to emit gluons into the gap can change. Thus, with a wider gap, it is not a surprise that $\cV_{\mi \pi}$ now has a visible effect.

What is surprising, at least to us, is that the effect from $\cV_{\mi \pi}$ has about the same magnitude as the effect from $\Delta\cH$ and $\Delta\cV_\mathrm{Re}$, but has the opposite sign. Thus when we add the effects together, we are almost back to the LC+ curve.

Now look at $f(\bar p_\LT)$ for $3 < y_{12} < 4$. For $\bar p_\LT < 200 \GeV$, the LC+ curve is quite precisely a straight line, but now with a larger slope: $A_1 \approx 0.61$, continuing the previous trend. 

The effect of $\Delta\cH$ and $\Delta\cV_\mathrm{Re}$ in the region $\bar p_\LT > 380 \GeV$ has now grown very slightly. The effect of $\cV_{\mi \pi}$ still approximately cancels the effect of $\Delta\cH$ and $\Delta\cV_\mathrm{Re}$.

There is a new effect that is now visible. Beyond $\bar p_\LT \approx 200 \GeV$, the LC+ curve is no longer a good fit to a straight line. Rather, it curves up slightly.  One could imagine that there is a $\log^2(\bar p_\LT/p_\LT^\mathrm{cut})$ term added to the exponent in Eq.~(\ref{eq:fexponential}). However, a fit to the results for $\bar p_\LT < 200 \GeV$ that includes a $\log^2(\bar p_\LT/p_\LT^\mathrm{cut})$ contribution still gives very nearly a straight line, which does not fit the results for $\bar p_\LT > 200 \GeV$. Thus we have a non-logarithmic large $\bar p_\LT$ effect. Such an effect is to be expected because of what we might call momentum starvation. If the two jets that define the gap have equal transverse momenta, then the c.m.\ energy of the colliding partons that could scatter to make these jets is $[\hat s]^{1/2} = \bar p_\LT \exp(y_{12}/2)$. For $\bar p_\LT = 400 \GeV$ and $y_{12} = 3.5$, this is $[\hat s]^{1/2} = 2.3 \TeV$. Since parton distribution functions fall with momentum fraction $x$, it is quite improbable to have a parton collision with this much $[\hat s]^{1/2}$. It is then more improbable to have an even larger $[\hat s]^{1/2}$ needed to radiate a gluon with enough transverse momentum to destroy the gap. Thus the gap fraction $f(\bar p_\LT)$ should be larger than it would be if this effect were ignored. Furthermore, this effect should become more pronounced as $y_{12}$ increases.

The fact that a momentum starvation effect is visible in the \textsc{Deductor} results indicates that momentum conservation is important in the calculation.

Most of the trends that we have observed for $y_{12} < 4$ continue for $4 < y_{12} < 5$ and $5 < y_{12} < 6$. 

For $\bar p_\LT < 200 \GeV$, the LC+ curves are still quite precisely straight lines. However, the slopes do not grow with $y_{12}$. For the regions $3 < y_{12} < 4$, $4 < y_{12} < 5$ and $5 < y_{12} < 6$ we have, respectively, $A_1 \approx 0.61$, $A_1 \approx 0.65$ and $A_1 \approx 0.56$. 

As we expect, the upward turn of the LC+ curve for $\bar p_\LT > 200 \GeV$ becomes more pronounced as $y_{12}$ increases.

The effect of including $\Delta\cH$ and $\Delta\cV_\mathrm{Re}$ grows as $y_{12}$ increases. In the region $\bar p_\LT > 300 \GeV$, This effect multiplies $f$ by a factor of about $\exp(- 0.41)$ for $4 < y_{12} < 5$ and $\exp(- 0.81)$ for $5 < y_{12} < 6$. 

The previous trend of an increasing effect from $\cV_{\mi\pi}$ does not continue. The effect from including $\cV_{\mi\pi}$ is to multiply $f$ in the region $\bar p_\LT > 300 \GeV$ by a factor of about $\exp(+ 0.14)$ for $3 < y_{12} < 4$, $\exp(+ 0.18)$ for $4 < y_{12} < 5$, and $\exp(+ 0.14)$ for $5 < y_{12} < 6$. Thus the $\cV_{\mi\pi}$ effect does not cancel the growing effect of $\Delta\cH$ and $\Delta\cV_\mathrm{Re}$. 

In principle, there should be contributions to $f$ proportional to $\cV_{\mi\pi}^2$ that contain an extra power of $\log(\bar p_\LT/p_\LT^\mathrm{cut})$, dubbed a ``superleading log'' \cite{Manchester2009, SuperLeading1, SuperLeading2, Manchester2011}. These contributions are surely present, but they are not large enough to be visible in the difference between the red and green curves in Fig.~\ref{fig:fR04}.

We are left with a net decrease in $f$ for $\bar p_\LT > 300 \GeV$ from color beyond the LC+ approximation by a factor of about $\exp(- 0.23)$ for $4 < y_{12} < 5$ and $\exp(- 0.61)$ for $5 < y_{12} < 6$.

\subsection{\label{sec:gap0207}Gap fraction dependence on $R$}

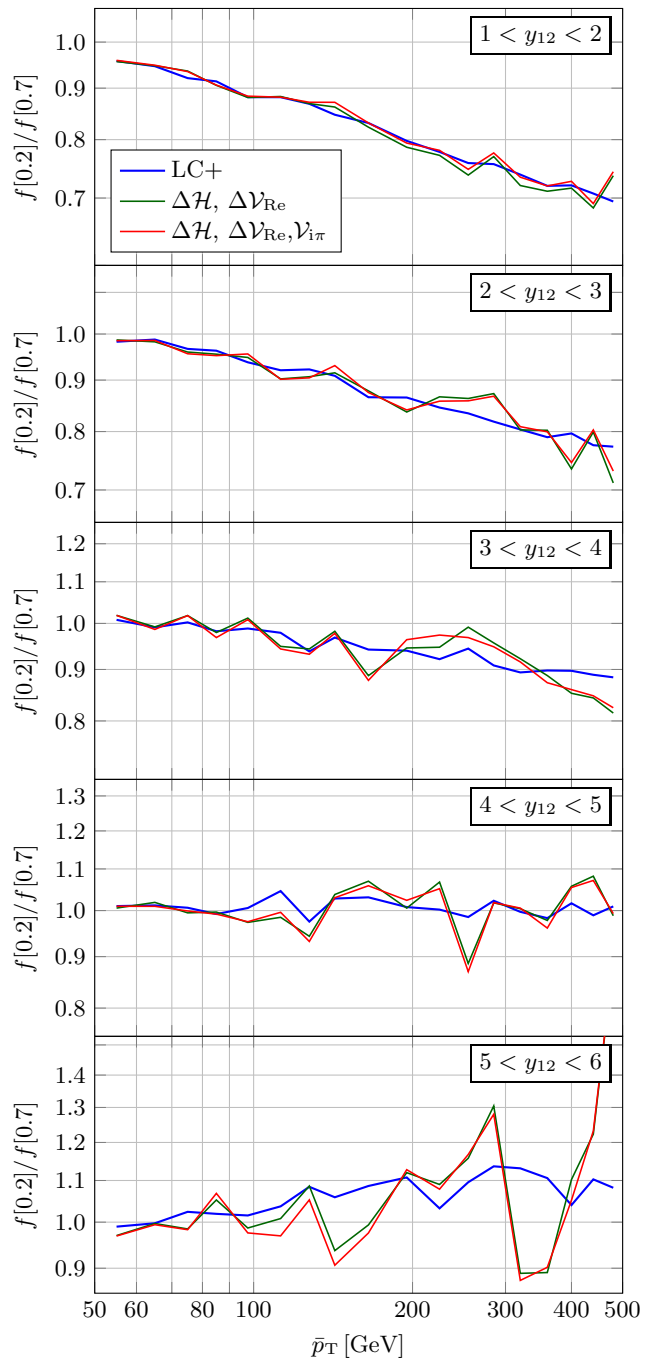
\begin{figure}
\begin{center}
\ifusefigs 

\begin{tikzpicture}
 \begin{groupplot}[
      group style={
          group size=1 by 5,
          vertical sep=0pt,
          x descriptions at=edge bottom},
          xlabel={$\bar p_\LT\,\mathrm{[GeV]}$},
          width=8.6cm,
    ]


\nextgroupplot[
ylabel={$f[0.2]/f[0.7]$},
height=5cm,
xmin=50, xmax=500,
ymin=0.6, ymax=1.08, 
xmode=log,
ymode=log,
ytick={0.6,0.7,0.8,0.9,1.0},
yticklabels={,0.7,0.8,0.9,1.0},
xtick={50,60,70,80,90,100,200,300,400,500},
  legend cell align=left,
  every axis legend/.append style={
  at={(0.03,0.05)},
  anchor = south west},
]

\pgfplotstableread{
55.     0.957675
65.     0.946352
75.     0.920901
85.     0.914051
97.5    0.881615
112.5   0.881574
127.5   0.868308
142.5   0.846891
165.    0.831062
195.    0.79776
225.    0.777903
255.    0.758366
285.    0.756565
320.    0.738591
360.    0.719803
400.    0.720593
440.    0.707273
480.    0.69448
}\ductLCpA

\addlegendentry{LC+}

\pgfplotstableread{
55.     0.956418
65.     0.947442
75.     0.935958
85.     0.905821
97.5    0.881004
112.5   0.882975
127.5   0.868757
142.5   0.861967
165.    0.823147
195.    0.78629
225.    0.7718
255.    0.737784
285.    0.769645
320.    0.720312
360.    0.710948
400.    0.716047
440.    0.684285
480.    0.736706
}\ductRealA

\addlegendentry{$\Delta\cH$, $\Delta \cV_\mathrm{Re}$}

\pgfplotstableread{
55.     0.959095
65.     0.948476
75.     0.934777
85.     0.906383
97.5    0.883711
112.5   0.881584
127.5   0.87143
142.5   0.871337
165.    0.831185
195.    0.793834
225.    0.780734
255.    0.747496
285.    0.776119
320.    0.734167
360.    0.719666
400.    0.727213
440.    0.691179
480.    0.743425
}\ductAllA

\addlegendentry{$\Delta\cH$, $\Delta \cV_\mathrm{Re}$,$\cV_{\mi\pi}$ }

\addplot [blue, thick] table [x={0},y={1}]{\ductLCpA};

\addplot [darkgreen, semithick] table [x={0},y={1}]{\ductRealA};

\addplot [red, semithick] table [x={0},y={1}]{\ductAllA};

\node at (350.0,1.08*0.94) {\fcolorbox{black}{white}{\small{$1 < y_{12} < 2$}}};


\nextgroupplot[
ylabel = {$f[0.2]/f[0.7]$},
height = 5 cm,
xmin=50, xmax=500,
ymin=0.65, ymax=1.17, 
xmode=log,
ymode=log,
ytick={0.6,0.7,0.8,0.9,1.0,1.1},
yticklabels={,0.7,0.8,0.9,1.0,},
xtick={50,60,70,80,90,100,200,300,400,500},
]

\pgfplotstableread{
55.     0.982529
65.     0.987466
75.     0.966595
85.     0.962606
97.5    0.937376
112.5   0.920454
127.5   0.92224
142.5   0.908819
165.    0.865392
195.    0.864813
225.    0.845186
255.    0.834065
285.    0.818364
320.    0.803871
360.    0.789795
400.    0.796686
440.    0.775546
480.    0.772925
}\ductLCpB

\pgfplotstableread{
55.     0.986589
65.     0.982283
75.     0.959649
85.     0.954953
97.5    0.947825
112.5   0.902445
127.5   0.90695
142.5   0.915075
165.    0.878075
195.    0.836584
225.    0.866187
255.    0.862921
285.    0.872625
320.    0.803003
360.    0.802359
400.    0.734455
440.    0.799335
480.    0.711419
}\ductRealB

\pgfplotstableread{
55.     0.984869
65.     0.985414
75.     0.955862
85.     0.951787
97.5    0.955452
112.5   0.901645
127.5   0.904611
142.5   0.93032
165.    0.874795
195.    0.840599
225.    0.857646
255.    0.858267
285.    0.867408
320.    0.809027
360.    0.799653
400.    0.745287
440.    0.803016
480.    0.730963
}\ductAllB

\addplot [blue, thick] table [x={0},y={1}]{\ductLCpB};

\addplot [darkgreen, semithick] table [x={0},y={1}]{\ductRealB};

\addplot [red, semithick] table [x={0},y={1}]{\ductAllB};

\node at (350.0,1.17*0.94) {\fcolorbox{black}{white}{\small{$2 < y_{12} < 3$}}};


\nextgroupplot[
ylabel = {$f[0.2]/f[0.7]$},
height = 5 cm,
xmin=50, xmax=500,
ymin=0.7, ymax=1.26, 
xmode=log,
ymode=log,
ytick={0.8,0.9,1.0,1.1,1.2},
yticklabels={0.8,0.9,1.0,1.1,1.2},
xtick={50,60,70,80,90,100,200,300,400,500},]

\pgfplotstableread{
55.     1.00825
65.     0.991041
75.     1.00238
85.     0.98225
97.5    0.988423
112.5   0.978785
127.5   0.937973
142.5   0.968239
165.    0.941948
195.    0.93961
225.    0.921458
255.    0.944047
285.    0.908336
320.    0.893954
360.    0.897903
400.    0.897275
440.    0.889154
480.    0.883972
}\ductLCpC

\pgfplotstableread{
55.     1.01862
65.     0.992113
75.     1.01809
85.     0.979439
97.5    1.01224
112.5   0.948805
127.5   0.943312
142.5   0.98219
165.    0.887435
195.    0.945354
225.    0.947
255.    0.991496
285.    0.956405
320.    0.922405
360.    0.887557
400.    0.852276
440.    0.843261
480.    0.814635
}\ductRealC

\pgfplotstableread{
55.     1.01858
65.     0.986343
75.     1.0181
85.     0.968103
97.5    1.00862
112.5   0.943208
127.5   0.931937
142.5   0.977593
165.    0.877826
195.    0.963515
225.    0.97369
255.    0.968161
285.    0.947922
320.    0.915686
360.    0.873125
400.    0.859506
440.    0.84737
480.    0.82458
}\ductAllC

\addplot [blue, thick] table [x={0},y={1}]{\ductLCpC};

\addplot [darkgreen, semithick] table [x={0},y={1}]{\ductRealC};

\addplot [red, semithick] table [x={0},y={1}]{\ductAllC};

\node at (350.0,1.26*0.94) {\fcolorbox{black}{white}{\small{$3 < y_{12} < 4$}}};


\nextgroupplot[
ylabel = {$f[0.2]/f[0.7]$},
height = 5 cm,
xmin=50, xmax=500,
ymin=0.75, ymax=1.35, 
xmode=log,
ymode=log,
ytick={0.8,0.9,1.0,1.1,1.2,1.3},
yticklabels={0.8,0.9,1.0,1.1,1.2,1.3},
xtick={50,60,70,80,90,100,200,300,400,500},]

\pgfplotstableread{
55.     1.01035
65.     1.01195
75.     1.00652
85.     0.992091
97.5    1.00588
112.5   1.04569
127.5   0.9751
142.5   1.028
165.    1.03079
195.    1.00802
225.    1.00223
255.    0.985315
285.    1.02274
320.    0.997106
360.    0.982928
400.    1.0168
440.    0.989289
480.    1.00961
}\ductLCpD

\pgfplotstableread{
55.     1.00572
65.     1.01891
75.     0.994822
85.     0.996631
97.5    0.973385
112.5   0.984731
127.5   0.942903
142.5   1.03751
165.    1.06954
195.    1.00569
225.    1.06744
255.    0.886065
285.    1.01783
320.    1.00425
360.    0.977697
400.    1.05723
440.    1.08193
480.    0.98828
}\ductRealD

\pgfplotstableread{
55.     1.00989
65.     1.00981
75.     0.999117
85.     0.991929
97.5    0.974859
112.5   0.996062
127.5   0.931759
142.5   1.02992
165.    1.05848
195.    1.02378
225.    1.05141
255.    0.86962
285.    1.01839
320.    1.00567
360.    0.96077
400.    1.05415
440.    1.07139
480.    0.99307
}\ductAllD

\addplot [blue, thick] table [x={0},y={1}]{\ductLCpD};

\addplot [darkgreen, semithick] table [x={0},y={1}]{\ductRealD};

\addplot [red, semithick] table [x={0},y={1}]{\ductAllD};

\node at (350.0,1.35*0.94) {\fcolorbox{black}{white}{\small{$4 < y_{12} < 5$}}};


\nextgroupplot[
ylabel={$f[0.2]/f[0.7]$},
height=5cm,
xmin=50, xmax=500,
ymin=0.85, ymax=1.53, 
xmode=log,
ymode=log,
ytick={0.9,1.0,1.1,1.2,1.3,1.4,1.5},
yticklabels={0.9,1.0,1.1,1.2,1.3,1.4,},
xtick={50,60,70,80,90,100,200,300,400,500},
xticklabels={50,60,,80,,100,200,300,400,500},]

\pgfplotstableread{
55.     0.989814
65.     0.99742
75.     1.02393
85.     1.01918
97.5    1.01542
112.5   1.03683
127.5   1.08424
142.5   1.05859
165.    1.08629
195.    1.10775
225.    1.03206
255.    1.09562
285.    1.13625
320.    1.13106
360.    1.10606
400.    1.03981
440.    1.10286
480.    1.08187
}\ductLCpE

\pgfplotstableread{
55.     0.969804
65.     0.996935
75.     0.984574
85.     1.05207
97.5    0.986725
112.5   1.00825
127.5   1.08625
142.5   0.93696
165.    0.993629
195.    1.11994
225.    1.09008
255.    1.15745
285.    1.30452
320.    0.889521
360.    0.891124
400.    1.10233
440.    1.22302
480.    1.81595
}\ductRealE

\pgfplotstableread{
55.     0.968776
65.     0.994173
75.     0.982911
85.     1.06807
97.5    0.97569
112.5   0.969105
127.5   1.05231
142.5   0.906201
165.    0.975508
195.    1.12758
225.    1.07834
255.    1.16744
285.    1.27992
320.    0.87526
360.    0.901852
400.    1.0514
440.    1.23205
480.    1.81486
}\ductAllE

\addplot [blue, thick] table [x={0},y={1}]{\ductLCpE};

\addplot [darkgreen, semithick] table [x={0},y={1}]{\ductRealE};

\addplot [red, semithick] table [x={0},y={1}]{\ductAllE};

\node at (350.0,1.53*0.94) {\fcolorbox{black}{white}{\small{$5 < y_{12} < 6$}}};
\end{groupplot}
\end{tikzpicture}

\else 
\NOTE{Figure fig:fRratio goes here.}
\fi

\end{center}
\caption{
Gap fraction ratio $f[R=0.2]/f[R=0.7]$ versus $y_{12}$ and $\bar p_\LT$.
}
\label{fig:fRratio}
\end{figure}

How does the choice of the cone size parameter $R$ affect the gap fraction? To find out, we carried out the previous calculation also for $R = 0.2$ and $R = 0.7$. We then divided $f[R = 0.2]$ by $f[R = 0.7]$ for each $y_{12}$ range and for each $\bar p_\LT$. The extent to which $f[R = 0.2]/f[R = 0.7]$ differs from 1 indicates the influence of $R$ on $f$. The results are plotted in Fig.~\ref{fig:fRratio}.

We see that when $y_{12}$ is not too large the gap fraction is smaller for $R = 0.2$ than it is for $R = 0.7$. This is easy to understand. There is a high probability to emit a gluon near the direction of one of the two leading jets that defines ends of the gap region. For a large jet radius $R$, this gluon is likely to form part of the jet. But for small $R$, this gluon can fall outside of the jet but inside the gap region, thus destroying the gap. That is, roughly collinear gluon radiation will decrease the gap fraction when $R$ is small. 

This effect of decreasing $f$ with decreasing $R$ diminishes as $y_{12}$ grows. This trend is easy to understand because for large $y_{12}$ there is a wide range for emission of a gluon that will destroy the gap, so that the range near the two jets that define the gap region is not so important. 

We note that for $5 < y_{12} < 6$ and $\bar p_\LT > 100 \GeV$, the gap fraction {\em increases} with decreasing $R$. This is not a large effect, but it is striking because it reverses the expected trend that we see for smaller $y_{12}$.

How do the extra color operators $\Delta \cH$ and $\Delta \cV$ affect the $R$ dependence of the gap fraction distribution? We see from Fig.~\ref{fig:fRratio} that there is no effect within the statistical errors. At the simplest level, this is easy to understand. The real emission operator $\Delta \cH$ reflects singularities for emission of soft gluons in directions that are not particularly close to the directions of existing partons. It does not contain collinear singularities. However, $R$ dependence for single emissions is connected with collinear singularities. Now, $R$ dependence could arise from a collinear emission followed by a wide angle soft emission, so we could see some $R$ dependence coming from $\Delta \cH$ and $\Delta \cV$. However, it is not a surprise that this dependence is small.

\begin{figure}[htb]
\begin{center}
\ifusefigs 

\begin{tikzpicture}
 \begin{groupplot}[
      group style={
          group size=1 by 5,
          vertical sep=0pt,
          x descriptions at=edge bottom},
          xlabel={$\bar p_\LT\,\mathrm{[GeV]}$},
          width=8.6cm,
    ]


\nextgroupplot[
ylabel={$f(\bar p_\LT)$},
height=5cm,
xmin=50, xmax=500,
ymin=0.1, ymax=1.0,
xmode=log,
ymode=log,
ytick={0.1,0.2,0.3,0.4,0.5,0.6,0.7,0.8,0.9,1.0},
yticklabels={,0.2,0.3,0.4,,0.6,,0.8,,1.0},
xtick={50,60,70,80,90,100,200,300,400,500},
  legend cell align=left,
  every axis legend/.append style={
  at={(0.03,0.05)},
  anchor = south west},
]

\pgfplotstableread{
55.     0.842327
65.     0.806814
75.     0.773002
85.     0.744255
97.5    0.714473
112.5   0.677057
127.5   0.654467
142.5   0.617054
165.    0.600561
195.    0.569743
225.    0.537639
255.    0.517033
285.    0.495017
320.    0.478954
360.    0.458889
400.    0.439664
440.    0.430567
480.    0.416685
}\NLOoneA

\pgfplotstableread{
55.     0.840698
65.     0.804562
75.     0.770609
85.     0.741827
97.5    0.711813
112.5   0.675196
127.5   0.651415
142.5   0.61644
165.    0.596952
195.    0.565244
225.    0.533527
255.    0.512278
285.    0.490294
320.    0.473245
360.    0.453342
400.    0.434473
440.    0.424364
480.    0.410838
}\NLOtwoA

\addlegendentry{NLO $\mu = 2 \bar p_T$}

\pgfplotstableread{
55.     0.842044
65.     0.80594
75.     0.772387
85.     0.743936
97.5    0.714108
112.5   0.678445
127.5   0.654104
142.5   0.621197
165.    0.599819
195.    0.567789
225.    0.53664
255.    0.515103
285.    0.493337
320.    0.475724
360.    0.456069
400.    0.437593
440.    0.426819
480.    0.413622
}\NLOfourA

\pgfplotstableread{
55.     0.840518
65.     0.798913
75.     0.763499
85.     0.734
97.5    0.69579
112.5   0.661739
127.5   0.638247
142.5   0.616068
165.    0.586371
195.    0.5526
225.    0.526907
255.    0.498009
285.    0.487869
320.    0.46635
360.    0.453832
400.    0.437371
440.    0.425733
480.    0.417114
}\ductLCpA

\addlegendentry{LC+}

\pgfplotstableread{
55.     0.842406
65.     0.799189
75.     0.76163
85.     0.73333
97.5    0.695443
112.5   0.65584
127.5   0.63153
142.5   0.618572
165.    0.584554
195.    0.537051
225.    0.525483
255.    0.490918
285.    0.505572
320.    0.471853
360.    0.448453
400.    0.437108
440.    0.414043
480.    0.42033
}\ductRealA

\addlegendentry{$\Delta\cH$, $\Delta \cV_\mathrm{Re}$}

\pgfplotstableread{
55.     0.84419
65.     0.804257
75.     0.768508
85.     0.740681
97.5    0.703389
112.5   0.667508
127.5   0.634201
142.5   0.634772
165.    0.603499
195.    0.552161
225.    0.545113
255.    0.508077
285.    0.519732
320.    0.48743
360.    0.463172
400.    0.453249
440.    0.427985
480.    0.436254
}\ductAllA

\addlegendentry{$\Delta\cH$, $\Delta \cV_\mathrm{Re}$,$\cV_{\mi\pi}$ }

\addplot [name path=pluserror, draw=none, forget plot]
 table [x={0},y={1}]{\NLOoneA};

\addplot [black, semithick] table [x={0},y={1}]{\NLOtwoA};

\addplot [name path=minuserror, draw=none, forget plot]
 table [x={0},y={1}]{\NLOfourA};

\addplot[forget plot, yellow] fill between[on layer={},of=pluserror and minuserror];

\addplot [blue, semithick] table [x={0},y={1}]{\ductLCpA};

\addplot [darkgreen, semithick] table [x={0},y={1}]{\ductRealA};

\addplot [red, semithick] table [x={0},y={1}]{\ductAllA};

\node at (350.0,0.8) {\fcolorbox{black}{white}{\small{$1 < y_{12} < 2$}}};


\nextgroupplot[
ylabel = $f(\bar p_\LT)$,
height=5cm,
xmin=50, xmax=500,
ymin=0.1, ymax=1.0,
xmode=log,
ymode=log,
ytick={0.1,0.2,0.3,0.4,0.5,0.6,0.7,0.8,0.9,1.0},
yticklabels={,0.2,0.3,0.4,,0.6,,0.8,,1.0},
xtick={50,60,70,80,90,100,200,300,400,500},
]

\pgfplotstableread{
55.     0.77101
65.     0.702713
75.     0.661774
85.     0.634334
97.5    0.584424
112.5   0.551607
127.5   0.529008
142.5   0.492048
165.    0.46294
195.    0.437174
225.    0.408797
255.    0.386243
285.    0.364834
320.    0.350676
360.    0.34262
400.    0.321176
440.    0.316689
480.    0.307278
}\NLOoneB

\pgfplotstableread{
55.     0.759844
65.     0.69267
75.     0.648861
85.     0.617763
97.5    0.567762
112.5   0.530985
127.5   0.503845
142.5   0.466785
165.    0.432824
195.    0.401692
225.    0.369882
255.    0.344326
285.    0.320882
320.    0.303251
360.    0.291357
400.    0.2688
440.    0.261044
480.    0.250104
}\NLOtwoB

\pgfplotstableread{
55.     0.756591
65.     0.691174
75.     0.646111
85.     0.612916
97.5    0.563575
112.5   0.524662
127.5   0.494703
142.5   0.458019
165.    0.421332
195.    0.386865
225.    0.353036
255.    0.325831
285.    0.301219
320.    0.281359
360.    0.266875
400.    0.243794
440.    0.23389
480.    0.222005
}\NLOfourB

\pgfplotstableread{
55.     0.764878
65.     0.711129
75.     0.659845
85.     0.63029
97.5    0.577288
112.5   0.537826
127.5   0.501771
142.5   0.479551
165.    0.441214
195.    0.411196
225.    0.38228
255.    0.360394
285.    0.343351
320.    0.329854
360.    0.314727
400.    0.297306
440.    0.290362
480.    0.283321
}\ductLCpB

\pgfplotstableread{
55.     0.769368
65.     0.708563
75.     0.654098
85.     0.630535
97.5    0.57621
112.5   0.528456
127.5   0.494705
142.5   0.475869
165.    0.434734
195.    0.401549
225.    0.367705
255.    0.354799
285.    0.340477
320.    0.307931
360.    0.312532
400.    0.27857
440.    0.265303
480.    0.244131
}\ductRealB

\pgfplotstableread{
55.     0.775029
65.     0.716019
75.     0.66219
85.     0.642467
97.5    0.590053
112.5   0.542435
127.5   0.514784
142.5   0.494494
165.    0.453119
195.    0.421344
225.    0.387949
255.    0.37101
285.    0.359428
320.    0.326792
360.    0.331543
400.    0.29846
440.    0.286284
480.    0.26211
}\ductAllB

\addplot [name path=pluserror, draw=none, forget plot]
 table [x={0},y={1}]{\NLOoneB};

\addplot [black, semithick] table [x={0},y={1}]{\NLOtwoB};

\addplot [name path=minuserror, draw=none, forget plot]
 table [x={0},y={1}]{\NLOfourB};

\addplot[forget plot, yellow] fill between[on layer={},of=pluserror and minuserror];

\addplot [blue, semithick] table [x={0},y={1}]{\ductLCpB};

\addplot [darkgreen, semithick] table [x={0},y={1}]{\ductRealB};

\addplot [red, semithick] table [x={0},y={1}]{\ductAllB};

\node at (350.0,0.8) {\fcolorbox{black}{white}{\small{$2 < y_{12} < 3$}}};


\nextgroupplot[
ylabel = $f(\bar p_\LT)$,
height=5cm,
xmin=50, xmax=500,
ymin=0.1, ymax=1.0,
xmode=log,
ymode=log,
ytick={0.1,0.2,0.3,0.4,0.5,0.6,0.7,0.8,0.9,1.0},
yticklabels={,0.2,0.3,0.4,,0.6,,0.8,,1.0},
xtick={50,60,70,80,90,100,200,300,400,500},]

\pgfplotstableread{
55.     0.702092
65.     0.635374
75.     0.578622
85.     0.561741
97.5    0.53696
112.5   0.491662
127.5   0.474883
142.5   0.431842
165.    0.445129
195.    0.409682
225.    0.412011
255.    0.414346
285.    0.403543
320.    0.406414
360.    0.387883
400.    0.4021
440.    0.401065
480.    0.400372
}\NLOoneC

\pgfplotstableread{
55.     0.671168
65.     0.599713
75.     0.538186
85.     0.508577
97.5    0.475455
112.5   0.423184
127.5   0.396366
142.5   0.352056
165.    0.347219
195.    0.303272
225.    0.292264
255.    0.283342
285.    0.266151
320.    0.259876
360.    0.237247
400.    0.241963
440.    0.236263
480.    0.230277
}\NLOtwoC

\pgfplotstableread{
55.     0.658237
65.     0.584791
75.     0.521275
85.     0.484737
97.5    0.446144
112.5   0.390462
127.5   0.357629
142.5   0.312843
165.    0.296751
195.    0.247932
225.    0.228417
255.    0.212415
285.    0.191292
320.    0.179229
360.    0.154228
400.    0.152838
440.    0.144221
480.    0.13522
}\NLOfourC

\pgfplotstableread{
55.     0.696003
65.     0.620445
75.     0.570868
85.     0.53878
97.5    0.491351
112.5   0.450772
127.5   0.42062
142.5   0.378167
165.    0.357102
195.    0.321557
225.    0.291665
255.    0.276664
285.    0.264924
320.    0.244382
360.    0.234631
400.    0.227397
440.    0.219717
480.    0.217836
}\ductLCpC

\pgfplotstableread{
55.     0.698916
65.     0.628799
75.     0.566458
85.     0.542248
97.5    0.495019
112.5   0.438461
127.5   0.389986
142.5   0.384994
165.    0.352162
195.    0.304028
225.    0.268094
255.    0.260305
285.    0.250634
320.    0.227774
360.    0.222138
400.    0.201605
440.    0.200021
480.    0.202425
}\ductRealC

\pgfplotstableread{
55.     0.705436
65.     0.640328
75.     0.583098
85.     0.555647
97.5    0.511811
112.5   0.458993
127.5   0.405698
142.5   0.404044
165.    0.369061
195.    0.318085
225.    0.290369
255.    0.276893
285.    0.26666
320.    0.245575
360.    0.237632
400.    0.220866
440.    0.218623
480.    0.222181
}\ductAllC

\addplot [name path=pluserror, draw=none, forget plot]
 table [x={0},y={1}]{\NLOoneC};

\addplot [black, semithick] table [x={0},y={1}]{\NLOtwoC};

\addplot [name path=minuserror, draw=none, forget plot]
 table [x={0},y={1}]{\NLOfourC};

\addplot[forget plot, yellow] fill between[on layer={},of=pluserror and minuserror];

\addplot [blue, semithick] table [x={0},y={1}]{\ductLCpC};

\addplot [darkgreen, semithick] table [x={0},y={1}]{\ductRealC};

\addplot [red, semithick] table [x={0},y={1}]{\ductAllC};

\node at (350.0,0.8) {\fcolorbox{black}{white}{\small{$3 < y_{12} < 4$}}};


\nextgroupplot[
ylabel = $f(\bar p_\LT)$,
height=5cm,
xmin=50, xmax=500,
ymin=0.1, ymax=1.0,
xmode=log,
ymode=log,
ytick={0.1,0.2,0.3,0.4,0.5,0.6,0.7,0.8,0.9,1.0},
yticklabels={,0.2,0.3,0.4,,0.6,,0.8,,1.0},
xtick={50,60,70,80,90,100,200,300,400,500},]

\pgfplotstableread{
55.     0.676157
65.     0.626603
75.     0.612571
85.     0.563822
97.5    0.580282
112.5   0.571976
127.5   0.536711
142.5   0.557406
165.    0.594024
195.    0.607247
225.    0.620543
255.    0.659128
285.    0.674905
320.    0.690396
360.    0.742258
400.    0.731733
440.    0.762835
480.    0.757772
}\NLOoneD

\pgfplotstableread{
55.     0.604113
65.     0.539548
75.     0.501931
85.     0.440634
97.5    0.433984
112.5   0.405814
127.5   0.360297
142.5   0.357166
165.    0.366292
195.    0.353515
225.    0.34675
255.    0.363473
285.    0.362257
320.    0.362899
360.    0.38905
400.    0.374681
440.    0.386184
480.    0.376867
}\NLOtwoD

\pgfplotstableread{
55.     0.570935
65.     0.497525
75.     0.447122
85.     0.380366
97.5    0.358898
112.5   0.319249
127.5   0.26854
142.5   0.251417
165.    0.24344
195.    0.215597
225.    0.196752
255.    0.199861
285.    0.189254
320.    0.181218
360.    0.192896
400.    0.17669
440.    0.178821
480.    0.168281
}\NLOfourD

\pgfplotstableread{
55.     0.614134
65.     0.553826
75.     0.517392
85.     0.479122
97.5    0.425718
112.5   0.401904
127.5   0.357247
142.5   0.327625
165.    0.299872
195.    0.275256
225.    0.253303
255.    0.23927
285.    0.224482
320.    0.217796
360.    0.212131
400.    0.207431
440.    0.204041
480.    0.207425
}\ductLCpD

\pgfplotstableread{
55.     0.630287
65.     0.567355
75.     0.543419
85.     0.48177
97.5    0.431408
112.5   0.412216
127.5   0.365044
142.5   0.332189
165.    0.289021
195.    0.254083
225.    0.21641
255.    0.189779
285.    0.196343
320.    0.177018
360.    0.185339
400.    0.143766
440.    0.145101
480.    0.171019
}\ductRealD

\pgfplotstableread{
55.     0.638322
65.     0.579003
75.     0.552482
85.     0.494077
97.5    0.442697
112.5   0.423841
127.5   0.379623
142.5   0.349576
165.    0.299791
195.    0.269318
225.    0.23228
255.    0.203213
285.    0.214877
320.    0.192609
360.    0.204034
400.    0.160387
440.    0.165567
480.    0.193562
}\ductAllD

\addplot [name path=pluserror, draw=none, forget plot]
 table [x={0},y={1}]{\NLOoneD};

\addplot [black, semithick] table [x={0},y={1}]{\NLOtwoD};

\addplot [name path=minuserror, draw=none, forget plot]
 table [x={0},y={1}]{\NLOfourD};

\addplot[forget plot, yellow] fill between[on layer={},of=pluserror and minuserror];

\addplot [blue, semithick] table [x={0},y={1}]{\ductLCpD};

\addplot [darkgreen, semithick] table [x={0},y={1}]{\ductRealD};

\addplot [red, semithick] table [x={0},y={1}]{\ductAllD};

\node at (350.0,0.8) {\fcolorbox{black}{white}{\small{$4 < y_{12} < 5$}}};


\nextgroupplot[
ylabel={$f(\bar p_\LT)$},
height=5cm,
xmin=50, xmax=500,
ymin=0.1, ymax=1.0,
xmode=log,
ymode=log,
ytick={0.1,0.2,0.3,0.4,0.5,0.6,0.7,0.8,0.9,1.0},
yticklabels={0.1,0.2,0.3,0.4,,0.6,,0.8,,1.0},
xtick={50,60,70,80,90,100,200,300,400,500},
xticklabels={50,60,,80,,100,200,300,400,500},]

\pgfplotstableread{
55.     0.758947
65.     0.639748
75.     0.769623
85.     0.687718
97.5    0.709152
112.5   0.799874
127.5   0.796801
142.5   0.877911
165.    0.936597
195.    1.05007
225.    1.1089
255.    1.17808
285.    1.25565
320.    1.28553
360.    1.39589
400.    1.40008
440.    1.59402
480.    1.70057
}\NLOoneE

\pgfplotstableread{
55.     0.606291
65.     0.480691
75.     0.531789
85.     0.459664
97.5    0.430296
112.5   0.467016
127.5   0.444785
142.5   0.480866
165.    0.493023
195.    0.550996
225.    0.56589
255.    0.596424
285.    0.630225
320.    0.631096
360.    0.683791
400.    0.663294
440.    0.757863
480.    0.805551
}\NLOtwoE

\pgfplotstableread{
55.     0.531975
65.     0.40498
75.     0.413892
85.     0.343679
97.5    0.29083
112.5   0.295935
127.5   0.262029
142.5   0.272844
165.    0.259872
195.    0.285205
225.    0.278527
255.    0.28879
285.    0.302035
320.    0.291739
360.    0.319444
400.    0.293821
440.    0.350952
480.    0.379889
}\NLOfourE

\pgfplotstableread{
55.     0.534401
65.     0.493032
75.     0.449917
85.     0.434367
97.5    0.391802
112.5   0.360165
127.5   0.337951
142.5   0.307465
165.    0.291556
195.    0.269378
225.    0.246162
255.    0.255607
285.    0.237343
320.    0.2448
360.    0.254609
400.    0.246019
440.    0.256981
480.    0.282283
}\ductLCpE

\pgfplotstableread{
55.     0.546572
65.     0.489224
75.     0.445746
85.     0.421993
97.5    0.419914
112.5   0.372338
127.5   0.33981
142.5   0.250403
165.    0.261904
195.    0.241868
225.    0.20227
255.    0.192121
285.    0.190201
320.    0.15512
360.    0.185308
400.    0.151707
440.    0.131917
480.    0.168073
}\ductRealE

\pgfplotstableread{
55.     0.552585
65.     0.493972
75.     0.452348
85.     0.424831
97.5    0.422773
112.5   0.367604
127.5   0.349166
142.5   0.25376
165.    0.271546
195.    0.253071
225.    0.209343
255.    0.210271
285.    0.199244
320.    0.172916
360.    0.202647
400.    0.174901
440.    0.149248
480.    0.194381
}\ductAllE

\addplot [name path=pluserror, draw=none, forget plot]
 table [x={0},y={1}]{\NLOoneE};

\addplot [black, semithick] table [x={0},y={1}]{\NLOtwoE};

\addplot [name path=minuserror, draw=none, forget plot]
 table [x={0},y={1}]{\NLOfourE};

\addplot[forget plot, yellow] fill between[on layer={},of=pluserror and minuserror];

\addplot [blue, semithick] table [x={0},y={1}]{\ductLCpE};

\addplot [darkgreen, semithick] table [x={0},y={1}]{\ductRealE};

\addplot [red, semithick] table [x={0},y={1}]{\ductAllE};

\node at (350.0,0.8) {\fcolorbox{black}{white}{\small{$5 < y_{12} < 6$}}};
\end{groupplot}
\end{tikzpicture}

\else 
\NOTE{Figure fig:NLO goes here.}
\fi

\end{center}
\caption{
Gap fraction calculated perturbatively to NLO using $\mu_\LR = \mu_\LF = 2 \bar p_\LT$, with an error band for $\bar p_\LT < \mu_\LR, \mu_\LF < 4 \bar p_\LT$. The \textsc{Deductor} curves from Fig.~\ref{fig:fR04} are also shown. 
}
\label{fig:fNLOR04}
\end{figure}
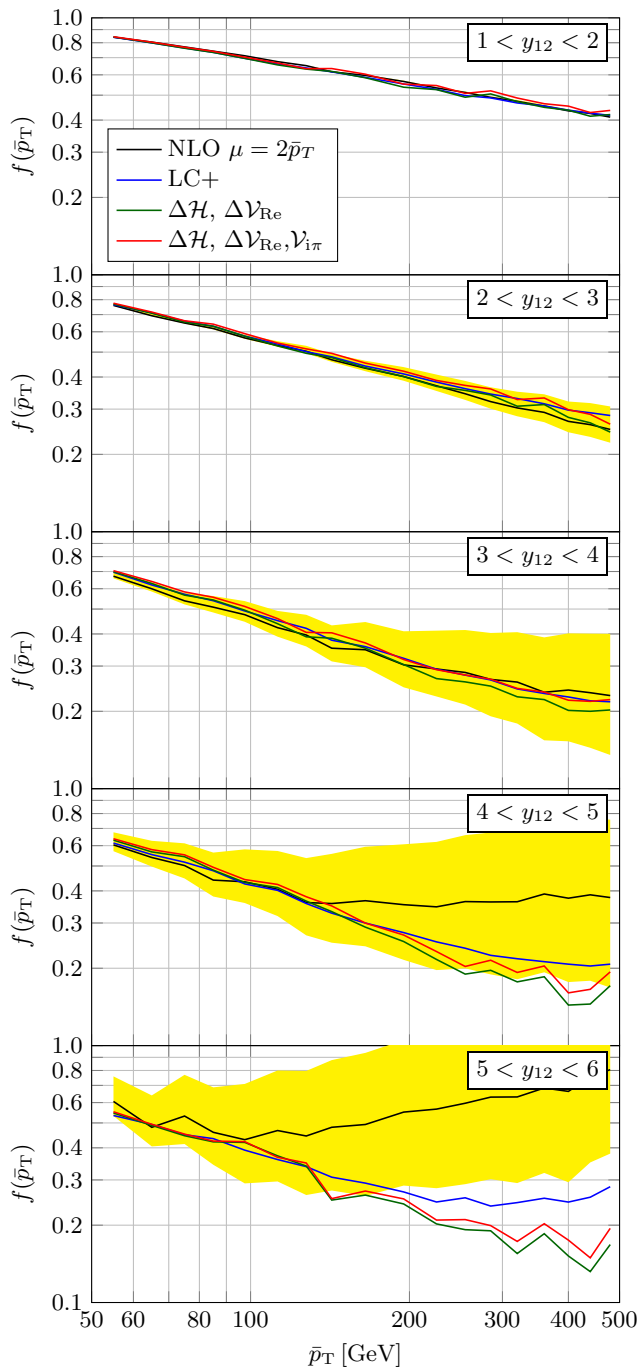

\subsection{\label{sec:gappert}Gap fraction in perturbation theory}

We can calculate the gap fraction in fixed order perturbation theory instead of using a parton shower. We write the gap fraction in the form of Eq.~(\ref{eq:fgapmod}). Here $d\sigma(\mathrm{total})/[d\bar p_\LT\,dy_{12}]$ is the cross section to produce at least two $R = 0.4$ jets in the rapidity window $-4.4 < y < 4.4$ such that the two jets in the rapidity window with the largest $P_\LT$ satisfy $\bar p_{\LT} = (P_{\LT,1} + P_{\LT,2})/2$ and $y_{12} = |y_1 - y_2|$. This is an infrared safe jet cross section for which the lowest order contribution has two partons in the final state. The cross section $d\sigma(\mathrm{no\ gap})/[d\bar p_\LT\,dy_{12}]$ is the cross section to produce at least three $R = 0.4$ jets in the rapidity window $-4.4 < y < 4.4$ such that the two jets in the rapidity window with the largest $P_\LT$ satisfy $\bar p_{\LT} = (P_{\LT,1} + P_{\LT,2})/2$ and $y_{12} = |y_1 - y_2|$ and such that there is a third jet with $\min(y_1,y_2) < y_3 < \max(y_1,y_2)$ and $P_{\LT,3} > p_{\LT}^{\rm cut} = 20 \GeV$. This is an infrared safe jet cross section for which the lowest order contribution has three partons in the final state. We calculate both $d\sigma(\mathrm{total})/[d\bar p_\LT\,dy_{12}]$ and $d\sigma(\mathrm{no\ gap})/[d\bar p_\LT\,dy_{12}]$ at NLO using \textsc{NLOJet++} \cite{NLOJet++}. For these perturbative calculations, our primary choice for the factorization and renormalization scales is $\mu_\LF = \mu_\LR = 2 \bar p_\LT$. We investigate the scale dependence by also using $\mu_\LF = \mu_\LR = \bar p_\LT$ and $\mu_\LF = \mu_\LR = 4 \bar p_\LT$.

In Ref.~\cite{NSThresholdII}, we calculated the gap fraction in this manner for $\sqrt{s} = 7 \TeV$. We found that, although there was substantial dependence on the scale choice for large $y_{12}$, the perturbative calculation with $\mu_\LF = \mu_\LR = 2 \bar p_\LT$ worked quite well. This was surprising to us because there are large logarithms that are not summed in the perturbative calculation. In this paper, we have chosen $\sqrt{s} = 13 \TeV$. Now there is a larger range available for gluon emissions.

Fig.~\ref{fig:fNLOR04}, we show the perturbative results for $\sqrt{s} = 13 \TeV$. For each range of $y_{12}$, we show a central curve with $\mu_\LF = \mu_\LR = 2 \bar p_\LT$. We show how the result varies for   $\bar p_\LT < \mu_\LF = \mu_\LR < 4 \bar p_\LT$ as a yellow error band. We also show the \textsc{Deductor} results from Fig.~\ref{fig:fR04}. We see that the perturbative results for $f$ agree with the \textsc{Deductor} results within about 20\% for $1 < y_{12} < 2$ and $2 < y_{12} < 3$. For $3 < y_{12} < 4$, the agreement between the two types of calculation is still good, but the scale variation error band on the perturbative calculation has grown substantially. For $4 < y_{12} < 5$ and $5 < y_{12} < 6$, the scale variation error band is so large that one can conclude that the NLO perturbative calculation is not reliable. Thus one needs either a parton shower calculation or an analytic summation of the large logarithms.

\subsection{Gap fraction in \textsc{Pythia}}

We can also compare the gap fraction results from \textsc{Deductor} with the analogous results from \textsc{Pythia} \cite{pythia}. Since \textsc{Pythia} is limited to the leading color approximation, we compare to \textsc{Deductor} with the LC+ approximation. Our previous investigations \cite{NSThresholdII} have indicated that non-perturbative effects are quite small for the gap fraction, so we have not included any non-perturbative effects in the results from \textsc{Deductor} in this paper.\footnote{\textsc{Deductor} itself includes only a parton shower based on perturbative splitting functions. However, on can, if desired, add an underlying event as contained in a non-perturbative model and then send the resulting partons to \textsc{Pythia} for hadronization.\cite{NSThresholdII}.} Accordingly, we have not  included the underlying event and hadronization in \textsc{Pythia}.  We used \textsc{Pythia} 8.423 with default settings except that we set $\as(M_Z^2) = 0.118$ in the shower and include the ``CMW'' factor in the $\as$ scale for shower splittings. In this way, we match the $\as$ settings used in the \textsc{Deductor} shower. Of course, \textsc{Pythia} and \textsc{Deductor} use very different algorithms to generate their parton showers. Of particular note are the inclusion of a summation of threshold logarithms in \textsc{Deductor} and the differences between \textsc{Pythia} and \textsc{Deductor} in splitting functions, shower ordering variable, and momentum mapping at each splitting. Thus we can expect only rough agreement between the programs. Nevertheless, it is of interest to see how much disagreement there is.

\begin{figure}
\begin{center}
\ifusefigs 

\begin{tikzpicture}
 \begin{groupplot}[
      group style={
          group size=1 by 5,
          vertical sep=0pt,
          x descriptions at=edge bottom},
          xlabel={$\bar p_\LT\,\mathrm{[GeV]}$},
          width=8.6cm,
    ]


\nextgroupplot[
ylabel={$f(\bar p_\LT)$},
height=5cm,
xmin=50, xmax=500,
ymin=0.1, ymax=1.0,
xmode=log,
ymode=log,
ytick={0.1,0.2,0.3,0.4,0.5,0.6,0.7,0.8,0.9,1.0},
yticklabels={,0.2,0.3,0.4,,0.6,,0.8,,1.0},
xtick={50,60,70,80,90,100,200,300,400,500},
  legend cell align=left,
  every axis legend/.append style={
  at={(0.03,0.05)},
  anchor = south west},
]

\pgfplotstableread{
55.     0.840518
65.     0.798913
75.     0.763499
85.     0.734
97.5    0.69579
112.5   0.661739
127.5   0.638247
142.5   0.616068
165.    0.586371
195.    0.5526
225.    0.526907
255.    0.498009
285.    0.487869
320.    0.46635
360.    0.453832
400.    0.437371
440.    0.425733
480.    0.417114
}\ductLCpA

\pgfplotstableread{
   55.0   0.85312
   65.0   0.80663
   75.0   0.77802
   85.0   0.75411
   97.5   0.71616
  112.5   0.68605
  127.5   0.66331
  142.5   0.64406
  165.0   0.61407
  195.0   0.58568
  225.0   0.56497
  255.0   0.54367
  285.0   0.52897
  320.0   0.51507
  360.0   0.50062
  400.0   0.48846
  440.0   0.47896
  480.0   0.47002
}\PythiaA

\addlegendentry{\textsc{Deductor} LC+}
\addlegendentry{\textsc{Pythia}}

\addplot [blue, semithick] table [x={0},y={1}]{\ductLCpA};
\addplot [red, semithick] table [x={0},y={1}]{\PythiaA};

\node at (350.0,0.8) {\fcolorbox{black}{white}{\small{$1 < y_{12} < 2$}}};


\nextgroupplot[
ylabel = $f(\bar p_\LT)$,
height = 5 cm,
xmin=50, xmax=500,
ymin=0.1, ymax=1.0,
xmode=log,
ymode=log,
ytick={0.1,0.2,0.3,0.4,0.5,0.6,0.7,0.8,0.9,1.0},
yticklabels={,0.2,0.3,0.4,,0.6,,0.8,,1.0},
xtick={50,60,70,80,90,100,200,300,400,500},
]

\pgfplotstableread{
55.     0.764878
65.     0.711129
75.     0.659845
85.     0.63029
97.5    0.577288
112.5   0.537826
127.5   0.501771
142.5   0.479551
165.    0.441214
195.    0.411196
225.    0.38228
255.    0.360394
285.    0.343351
320.    0.329854
360.    0.314727
400.    0.297306
440.    0.290362
480.    0.283321
}\ductLCpB

\pgfplotstableread{
   55.0   0.76062
   65.0   0.71162
   75.0   0.66312
   85.0   0.61853
   97.5   0.58668
  112.5   0.54761
  127.5   0.51651
  142.5   0.49278
  165.0   0.46203
  195.0   0.43112
  225.0   0.40971
  255.0   0.39045
  285.0   0.37609
  320.0   0.36395
  360.0   0.35086
  400.0   0.34131
  440.0   0.33401
  480.0   0.32815
}\PythiaB

\addplot [blue, semithick] table [x={0},y={1}]{\ductLCpB};
\addplot [red, semithick] table [x={0},y={1}]{\PythiaB};

\node at (350.0,0.8) {\fcolorbox{black}{white}{\small{$2 < y_{12} < 3$}}};


\nextgroupplot[
ylabel = $f(\bar p_\LT)$,
height = 5 cm,
xmin=50, xmax=500,
ymin=0.1, ymax=1.0,
xmode=log,
ymode=log,
ytick={0.1,0.2,0.3,0.4,0.5,0.6,0.7,0.8,0.9,1.0},
yticklabels={,0.2,0.3,0.4,,0.6,,0.8,,1.0},
xtick={50,60,70,80,90,100,200,300,400,500},]

\pgfplotstableread{
55.     0.696003
65.     0.620445
75.     0.570868
85.     0.53878
97.5    0.491351
112.5   0.450772
127.5   0.42062
142.5   0.378167
165.    0.357102
195.    0.321557
225.    0.291665
255.    0.276664
285.    0.264924
320.    0.244382
360.    0.234631
400.    0.227397
440.    0.219717
480.    0.217836
}\ductLCpC

\pgfplotstableread{
   55.0   0.67083
   65.0   0.60201
   75.0   0.54096
   85.0   0.49333
   97.5   0.45412
  112.5   0.41559
  127.5   0.39208
  142.5   0.37762
  165.0   0.34365
  195.0   0.31745
  225.0   0.29649
  255.0   0.27907
  285.0   0.27093
  320.0   0.26305
  360.0   0.25523
  400.0   0.25173
  440.0   0.24599
  480.0   0.24663
}\PythiaC

\addplot [blue, semithick] table [x={0},y={1}]{\ductLCpC};
\addplot [red, semithick] table [x={0},y={1}]{\PythiaC};

\node at (350.0,0.8) {\fcolorbox{black}{white}{\small{$3 < y_{12} < 4$}}};


\nextgroupplot[
ylabel = $f(\bar p_\LT)$,
height = 5 cm,
xmin=50, xmax=500,
ymin=0.1, ymax=1.0,
xmode=log,
ymode=log,
ytick={0.1,0.2,0.3,0.4,0.5,0.6,0.7,0.8,0.9,1.0},
yticklabels={,0.2,0.3,0.4,,0.6,,0.8,,1.0},
xtick={50,60,70,80,90,100,200,300,400,500},]

\pgfplotstableread{
55.     0.614134
65.     0.553826
75.     0.517392
85.     0.479122
97.5    0.425718
112.5   0.401904
127.5   0.357247
142.5   0.327625
165.    0.299872
195.    0.275256
225.    0.253303
255.    0.23927
285.    0.224482
320.    0.217796
360.    0.212131
400.    0.207431
440.    0.204041
480.    0.207425
}\ductLCpD

\pgfplotstableread{
   55.0   0.58366
   65.0   0.50447
   75.0   0.43394
   85.0   0.39817
   97.5   0.37308
  112.5   0.33499
  127.5   0.30380
  142.5   0.27729
  165.0   0.26297
  195.0   0.23958
  225.0   0.22890
  255.0   0.21870
  285.0   0.21554
  320.0   0.21102
  360.0   0.21617
  400.0   0.21561
  440.0   0.21824
  480.0   0.22288
}\PythiaD

\addplot [blue, semithick] table [x={0},y={1}]{\ductLCpD};
\addplot [red, semithick] table [x={0},y={1}]{\PythiaD};

\node at (350.0,0.8) {\fcolorbox{black}{white}{\small{$4 < y_{12} < 5$}}};


\nextgroupplot[
ylabel={$f(\bar p_\LT)$},
height=5cm,
xmin=50, xmax=500,
ymin=0.1, ymax=1.0,
xmode=log,
ymode=log,
ytick={0.1,0.2,0.3,0.4,0.5,0.6,0.7,0.8,0.9,1.0},
yticklabels={0.1,0.2,0.3,0.4,,0.6,,0.8,,1.0},
xtick={50,60,70,80,90,100,200,300,400,500},
xticklabels={50,60,,80,,100,200,300,400,500},]

\pgfplotstableread{
55.     0.534401
65.     0.493032
75.     0.449917
85.     0.434367
97.5    0.391802
112.5   0.360165
127.5   0.337951
142.5   0.307465
165.    0.291556
195.    0.269378
225.    0.246162
255.    0.255607
285.    0.237343
320.    0.2448
360.    0.254609
400.    0.246019
440.    0.256981
480.    0.282283
}\ductLCpE

\pgfplotstableread{
   55.0   0.51167
   65.0   0.46142
   75.0   0.34487
   85.0   0.31481
   97.5   0.30455
  112.5   0.28079
  127.5   0.22418
  142.5   0.20556
  165.0   0.21477
  195.0   0.20848
  225.0   0.20924
  255.0   0.19626
  285.0   0.20036
  320.0   0.20639
  360.0   0.21836
  400.0   0.21282
  440.0   0.27607
  480.0   0.25469
}\PythiaE

\addplot [blue, semithick] table [x={0},y={1}]{\ductLCpE};
\addplot [red, semithick] table [x={0},y={1}]{\PythiaE};

\node at (350.0,0.8) {\fcolorbox{black}{white}{\small{$5 < y_{12} < 6$}}};
\end{groupplot}
\end{tikzpicture}

\else 
\NOTE{Figure fig:fR04Pythia goes here.}
\fi

\end{center}
\caption{
Gap fraction $f$ for $R = 0.4$ versus $y_{12}$ and $\bar p_\LT$ for \textsc{Pythia} and for \textsc{Deductor} with the LC+ approximation.
}
\label{fig:fR04Pythia}
\end{figure}
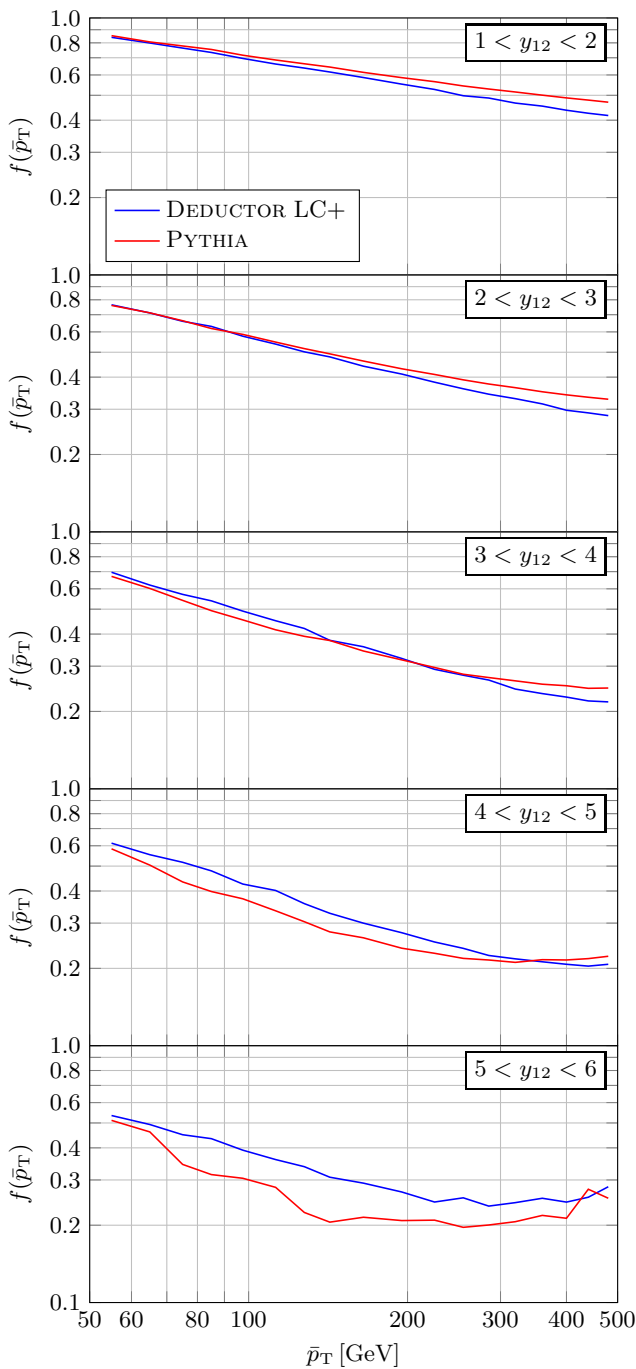

We exhibit the comparison in Fig.~\ref{fig:fR04Pythia}. We see that the differences between \textsc{Pythia} and \textsc{Deductor} with the LC+ approximation are fairly modest except at the largest values of $y_{12}$, for which it appears that \textsc{Pythia} produces more initial state radiation that can destroy the gap and thus make the gap fraction smaller.

\section{\label{sec:conclusions}Conclusions}

\textsc{Deductor} is a parton shower event generator whose primary purpose is to implement new theoretical developments in parton shower algorithms so as to improve the precision and predictive power of parton shower event generators. In a general framework \cite{NSAllOrder}, a parton shower is a solution of a renormalization group equation in which the generators of scale changes are, at order $\as$, operators called $\cH_I(t)$ and $\cV(t)$ in \textsc{Deductor}.

One of the important questions in this program is how to deal with color in the shower evolution. Color in parton showers has typically been treated in the leading color (LC) approximation. Previous versions of \textsc{Deductor} used the LC+ approximation, which is an improvement over the LC approximation but is nowhere near exact: $\cH_I(t)$ and $\cV(t)$ contain contributions $\Delta\cH(t)$ and $\Delta\cV(t)$ that are simply dropped in the LC+ approximation. The current version of \textsc{Deductor} \cite{NSMoreColor} allows one to include $\Delta\cH(t)$ and $\Delta\cV(t)$ perturbatively. That is, a limited number of powers of  $\Delta\cH(t)$ and $\Delta\cV(t)$ can be included in a calculation.

In Ref.~\cite{NSMoreColor}, we found that a calculation beyond the LC+ approximation could give numerical results for a physical cross section. We chose the one-jet-inclusive cross section and found an approximately 3\% effect from the extra color. In the present paper, we have asked whether \textsc{Deductor} with improved color can produce numerical results for more complicated physical cross sections and whether there are examples in which the effects of extra color are numerically at a level of $1/N_\Lc^2 \sim 10\%$ or higher.

We have chosen as our example the rapidity gap survival probability. This observable is of some practical significance because its study can help us to understand the effect in a search for new physics signals of vetoing against extra jets with transverse momenta greater than a cutoff $p_\LT^\mathrm{cut}$. It is also of special interest because it contains non-global logarithms, which are not simple to sum in an analytical approach.

We found in this study that the effects of extra color are substantial for soft gluon emissions when the rapidity difference, $y_{12}$, between the two leading jets and the average, $\bar p_\LT$, of their transverse momenta are large. We also found that the contribution of $\mi\pi$ terms in $\Delta \cV(t)$ saturates and does not grow significantly with the rapidity separation. It is interesting that these effects work into the opposite directions. The wide angle soft gluon emissions decrease the survival rate while the $\mi\pi$ terms increase it. Finally, we found that kinematic effects that result from exactly conserving momentum in the parton shower are numerically important.

Since we included the $\Delta\cH(t)$ and $\Delta\cV(t)$ only perturbatively, one can ask how many powers of these operators we need to make stable predictions. We tested this and we found that, for the gap fraction, the result is rather stable after two insertions of the soft correction operators. This finding is important because we cannot actually include many powers of $\Delta\cH(t)$ and $\Delta\cV(t)$. First, we cannot simply work to all orders in these operators. The dimension of the color space grows roughly as $(N!)^2$ with the  number of partons. In a typical shower calculation the averaged number of partons is 20-30. It is clear that there is no hope to deal with this problem exactly, so that one must use a perturbative approach. In the perturbative calculation we cannot include a very large power of the soft correction operators because the computer resource demand of the program gets out of control very quickly.  

Parton showers have their own systematic logic as operator renormalization group evolution \cite{NSAllOrder} in which, so far, we know only the order $\as$ contributions to the generators of scale changes. Using a parton shower to calculate an observable like the gap fraction $f$ has the effect of summing large logarithms. With a loose interpretation of what constitutes a logarithm, there are three sorts of large logarithms $L$ in $f$: $\log(\bar p_\LT/p_{\LT}^\mathrm{cut})$, $y_{12}$, and $\mi \pi$. Then in perturbation theory we have contributions $\as^n L^k$ with $k \le 2n$. One might hope that $\log f$ has an expansion with terms  $\as^n L^k$ with $k \le n+1$ in both full QCD and in an all orders parton shower and that a leading order parton shower gets the $\as L^2$ and $\as L$ contributions to $\log f$ correctly. However, a proof of this conjecture would not be easy and lies beyond the scope of this paper.

The calculation presented here does not include matching to NLO perturbation theory. In its current version, \textsc{Deductor} starts with the color density matrix for $2 \to 2$ QCD scattering at order $\as^2$. It then applies an operator $\cU_\cV$ that sums threshold logarithms, as described in Ref.~\cite{NSThresholdII}. After that, it applies the operator $\cU$ that generates parton splittings and thus a parton shower. Imagine expanding the combined operator $\cU\,\cU_\cV$ in powers of $\as$. The first term is simply the unit operator, but then there are terms proportional to $\as^1$ and higher powers of $\as$. The term proportional to $\as^1$, multiplying the $\as^2$ hard scattering color density matrix, gives an approximation to the order $\as^3$ density matrix. It is an approximation because the the operators $\cU$ and $\cU_\cV$ are based on soft and collinear limits. As explained in Ref.~\cite{NSAllOrder}, it is possible to include the complete order $\as^3$ color density matrix while correcting for the $\as^3$ contributions provided by $\cU$ and $\cU_\cV$. The procedure for this matching is straightforward, although it is more complex than the procedure needed to match cross sections because the object that needs matching is the color density matrix rather than just its trace over color. With NLO matching, calculations like the gap fraction calculation presented here would be more accurate and less sensitive to scale parameter choices such as the choice of the hardness scale at which the shower starts. We hope to add NLO matching in a future version of \textsc{Deductor}.

\begin{acknowledgments}
This work was supported in part by the United States Department of Energy under grant DE-SC0011640. This work benefited from access to the University of Oregon high performance computer, Talapas, and from access to the DESY Theory Group computer cluster.
\end{acknowledgments}



\end{document}